\documentclass[twocolumn,secnumarabic,amssymb, nobibnotes, aps, prd]{revtex4-2}

\usepackage{color}
\usepackage{graphicx}
\usepackage{subfigure}
\usepackage{amsmath}
\usepackage{dcolumn}
\usepackage{bm}

\begin{document}

\title{Designing a Concentrated High-Efficiency Thermionic Solar Cell Enabled by Graphene Collector}%

\author{Xin Zhang}
\affiliation{School of Science, Jiangnan University, Wuxi 214122, China}

\author{Cong Hu}
\author{Xiaohang Chen}
\author{Jincan Chen}
\email{Corresponding author: jcchen@xmu.edu.cn}
\affiliation{Department of Physics, Xiamen University, Xiamen 361005, China}

\author{Lay Kee Ang}
\author{Yee Sin Ang} 
\email{yeesin_ang@sutd.edu.sg}
\affiliation{Science, Math and Technology, Singapore University of Technology and Design, Singapore 487372}

\date{\today}%

\begin{abstract}

We propose a concentrated thermionic emission solar cell design, which demonstrates a high solar-to-electricity energy conversion efficiency larger than 10\% under 600 sun, by harnessing the exceptional electrical, thermal and radiative properties of the graphene as a collector electrode. By constructing an analytical model that explicitly takes into account the non-Richardson behavior of the thermionic emission current from graphene, space charge effect in vacuum gap, and the various irreversible energy losses within the subcomponents, we perform a detailed characterization on the conversion efficiency limit and electrical power output characteristics of the proposed system. We systematically model and compare the energy conversion efficiency of various configurations of graphene-graphene and graphene-diamond and diamond-diamond thermionic emitter, and show that utilizing diamond films as an emitter and graphene as a collector offers the highest maximum efficiency, thus revealing the important role of graphene in achieving high-performance thermionic emission solar cell. A maximum efficiency of 12.8\% under 800 sun has been revealed, which is significantly higher than several existing solid-state solar cell designs, such as the solar-driven thermoelectric and thermophotovoltaic converters. Our work thus opens up new avenues to advance the efficiency limit of thermionic solar energy conversion and the development of next-generation novel-nanomaterial-based solar energy harvesting technology.
\end{abstract}

\maketitle

\section{Introduction}
Solar energy, an abundant and clean energy source, is expected to play an essential role in easing the tense situation of fossil energy supply, optimizing the energy structure and reducing the environment pollution due to energy production and consumption \cite{green2017energy,gong2019advances,kamide2019heat}. Nowadays, photovoltaic (PV) cells \cite{rau2017efficiency}, photochemical approach \cite{guo2018photocorrosion}, and solar thermal power conversion \cite{chester2011design} are three main techniques for harvesting solar energy. Within a PV energy conversion process, photons with energies above the semiconductor bandgap excite an electron-hole pairs, which subsequently diffuse into charge-selective electrodes to form an electrical current. The photochemical process converts solar energy to storable chemical fuels, such as hydrogen. Solar thermal conversion, in which photons are converted into thermal energy by solar-thermal absorbers for subsequent electrical power generation, are widely used in industrial production and residential facilities, such as the mechanical heat engines in large-scale power plants and the solar water heaters in household heat supply system \cite{hussain2018advances,kraemer2011high}. The major advantages of the last approach are the exploitation of the entire solar spectrum and therefore it enables higher conversion efficiencies. More importantly, it is environmentally friendly with almost the smallest carbon footprint among all the energy harvesting approaches \cite{lin2020structured}.

Based on thermionic emission, the thermionic energy converter (TIEC) is an emerging heat-to-electric power generation system, with its performance ultimately limited by the Carnot efficiency \cite{khalid2016review,zebarjadi2017solid,zhang2017parametric}. Operated at relatively high temperatures, typically over 1400 K, electrons with energies larger than the emitter surface work function are thermally excited to escape from the emitter surface. When the system is connected to an external load, these emitted electrons traverse across the vacuum gap and are subsequently absorbed by the collector maintained at a lower temperature, thus generating an electrical power through the system. Compared with current technologies for heat-to-electricity conversion, such as the thermoelectric generator (TEG) \cite{kaur2019thermoelectric,kraemer2016concentrating} and thermophotovoltaics (TPV) \cite{bermel2010design,bhatt2020high,svetovoy2014graphene}, the TIEC operating at higher temperatures possesses substantially a higher theoretical efficiency and is thus particularly well suited for concentrated solar thermal-harvesting system and for high-grade waste recovery. For TEGs and TPVs, the high-temperature operation is plagued by multiple fundamental device limitations such as maintaining large temperature gradients in TEGs while minimizing the detrimental heat back-flow, and mitigating the large dark saturation currents in TPVs. In this aspect, the TIEC, which inherently operates at high-temperature regime where thermionic emission is profound, can significantly have more advantageous than the TEG and TPV systems.  

Despite being an appealing energy conversion technology that has been widely investigated for both industrial and aerospace applications, the demonstrated solar conversion efficiencies of TIECs are still typically below 10\% \cite{yuan2017back}. Such a low efficiency originates from the high work functions of the electron emitter and collector. For the routinely used tungsten and graphene, the rather high work functions of roughly 4.5 eV forms a major obstacle in obtaining high-power generation and conversion efficiency \cite{gubbels1989wf6,liang2015electron}. Such materials require extremely high temperatures to enable substantial electron emission current, which hampers the widespread applications for solar concentrators owing to the predominant radiative losses that seriously reduces their performance. Another major challenge is the inevitable presence of the space charge effect in practical vacuum thermionic devices \cite{smith2009theory,su2014space}. When the accumulation of space charge within the inter-electrode vacuum gap forms a potential barrier that suppress the electron emission, thus leading to a much reduced electrical current and power output. A comprehensive understanding of how these effects influence can quantitatively impacts the energy conversion efficiency of TIECs, and an optimal design strategy for achieving the efficiency larger than 10\% remains largely unanswered thus far. 

In this work, we propose a new design of high-efficiency solid-state concentrated thermionic solar cell (CTSC), enabled by a spectrally selective receiver and a graphene-based TIEC (Fig. 1). We design a system prototype with a parallel-plate geometry by incorporating advanced functional materials, such as thermally resilient hafnium carbide (HfC) based solar receiver, diamond films acting as thermionic emitter, and graphene as an electron collector. Importantly, we construct a comprehensive physics-based TIEC model by employing experimental parameters to obtain a realistic description of the TIEC system performance. Our mode suggests that graphene is a desirable material choice for designing high-performance collector, when compared to conventional bulk metal, due to the following facts: (i) Graphene quenches the reversed electron flux from the collector because of its low electron emission current density as limited by the vanishing density of states near the Dirac point, thus increasing the net electrical current generation; (ii) The exceptional physical properties, such as ultrahigh electrical mobility, low emissivity avoids heat radiation loss (2.3\%), and exhibits superior thermal stability (up till $4000$ K), thus making graphene a robust and low-loss collector; and (iii) The work function of graphene can be reduced to 1 eV by a combination of electrostatic gating and alkali metal deposition \cite{yuan2015engineering}, thus resulting in a larger output voltages. Because of these valuable advantages, back-gated graphene serves a promising collector material for boosting the power generation and conversion efficiency in TIECs, thus playing an important role in enhancing the performance of CTSC proposed in this work. 
It should be noted that although an idealized CTSC based on a graphene-based emitter has been predicted to yield a maximum efficiency of 21\% under 800 suns in previous work \cite{zhang2018graphene}, important physical effects, such as the space-charge effect, the non-Richardson thermionic emission behavior of graphene and other physical properties and the temperature dependence of the electrode materials, are not included in the prior model. The model presented in this work thus provide an important theoretical step-forward for the computational design of CTSC under more realistic operating conditions.

Particularly, our model takes into account two important, yet often missing in prior literature, effects: (i) Based on a full-band electronic band structure model of graphene, a generalized analytical model of space-charge limited thermionic emission current across the graphene/vacuum interface is developed and is incorporated in the modeling; and (ii) We further include the variation of the electrode temperatures by performing an energy balance analysis between the two electrodes. To understand the importance of graphene in the design of CTSC, we systematically study four different emitter-collector configurations, namely: (i) diamond-graphene, (ii) diamond-diamond, (iii) graphene-graphene, and (iv) graphene-diamond configurations. Our model reveals that the diamond-graphene emitter-collector configuration exhibits the largest solar-to-electricity conversion efficiency limit of $>$10\%, the largest value reported thus far for TIEC-based systems (see Table 1 below). Our findings demonstrate the feasibility of CTSC in advancing the performance limits of TIEC-based systems, and the important role of graphene as a thermionic anode material in achieving high-efficiency energy conversion. Our findings shed new light on the optimum design and the technological implementation of graphene-based solar thermal conversion devices for enriching the future clean energy landscape beyond fossil fuel.

\section{System Design and Modelling}

\subsection{Design principle of concentrated thermionic solar cells (CTSCs) using graphene as collector}

The concept of the proposed CTSC is composed of a radiation absorber and a TIEC unit, as depicted in Fig. 1(a). A transparent window allows the concentrated sunlight to impinge on the absorber, which is surface-textured for enhancing solar absorption and pursues minimum losses. The absorber harnesses the concentrated solar spectrum and transforms them into thermal energy. The TIEC stage exploits the absorbed heat, operating at the temperature range of 1000-1500 K, thus achieving thermal-to-electric energy conversion via thermionic emission. Simultaneously, the low-grade heat ($<$ 800 K) rejected from the collector is released to the environment via thermal radiation and conduction. In order to follow closely the absorber temperature, the emitter composed of a suitable low-work-function material is deposited on the absorber inner surface. A collector with a work function lower than that of the emitter establishes an output voltage across an external load. The whole system should work under a vacuum environment in order to ensure a sufficiently long mean free path for the emitted electrons and to avoid chemical-physical degradation of active materials when operated at high temperatures.

The material selection of the various subcomponents described above is as follows. HfC is selected as the absorber material in our design in order to realize the spectral selective cut-off absorber at high operating temperatures \cite{sani2012spectrally}. HfC possesses the high melting point (4153 K) and high electric conductivity ($4\times10^{-5}$ $\Omega$cm at room temperature), which are particularly advantageous in high-temperature operation and in reducing the series resistance and to avoid the bottlenecks for the refilling of emitted electrons. Excellent thermal conductivity (0.4 Wcm$^{-1}$K$^{-1}$ at 1773 K) enables the absorbed thermal energy to be transported efficiently to the emitter surface, without producing a wide temperature gradient between the light-receiving and electron-emitting surfaces. More importantly, HfC also has an intrinsic spectral selectivity, with high reflectance in the medium and far IR ranges, and low solar reflectance \cite{hans2018hafnium}.

The material selected for the emitter is N-doped H-terminated diamond films which have excellent electric and thermal properties \cite{koeck2012substrate}. Combining the advantages of negative electron affinity and Fermi level upshift through N-type doping, N-doped diamond films lower the effective work function to 1.7 eV and mitigate the space charge effect, thus improving the thermally driven electron emission capability \cite{koeck2011enhanced}. In addition, diamond films possess the highest thermal conductivity among the solids (ranging from 10 to 20 Wcm$^{-1}$K$^{-1}$), which is an essential property for closing to the receiver temperature with minimum thermal losses \cite{nemanich2014cvd}. For maximizing the output power, a collector with work function substantially lower than N-doped diamond films is necessary to obtain a sizable output voltage. For this purpose, monolayer graphene grown on copper foil by chemical vapor deposition and transferred onto Hafnium oxide (HfO$_2$) and silicon substrate for electrostatic gate-tuning capability is selected as the collector material \cite{yuan2017back}. Here, graphene serves as a practical collector material because of its exceptional physical properties such as ultrahigh electrical mobility, gate-tunable Fermi level, and superior thermal conductivity \cite{novoselov2005two,mortazavi2017ultra,zamani2018graphene}. Recently, an ultralow work function of 1.01 eV is obtained in an electrostatically back-gated graphene monolayer with Cs/O surface coating \cite{yuan2015engineering}, which has been successfully incorporated as a collector in a TIEC prototype. Electrostatic gating of graphene via a 20 nm HfO$_2$ dielectric layer allows the graphene work function to be dynamically tuned by 0.63 eV \cite{yuan2017back}. These superior physical properties of graphene justify its role as a low work function collector in our CTSC design.

The inter-electrode vacuum gap of the TIEC is chosen in the micrometer regime \cite{hishinuma2001refrigeration} because an overly narrow interelectrode gaps can lead to the collector overheating caused by near-filed heat transfer, whereas an overly large vacuum gap is dominated by the space charge effect which can negatively impact the performance of the thermionic emitter. Here we set the vacuum gap as 2 $\mu$m, in order to weaken the space charge effect and optimize other parameters such as emitter temperature to achieve high energy conversion efficiency.

\begin{figure}
\centering
\includegraphics[height=4.5cm]{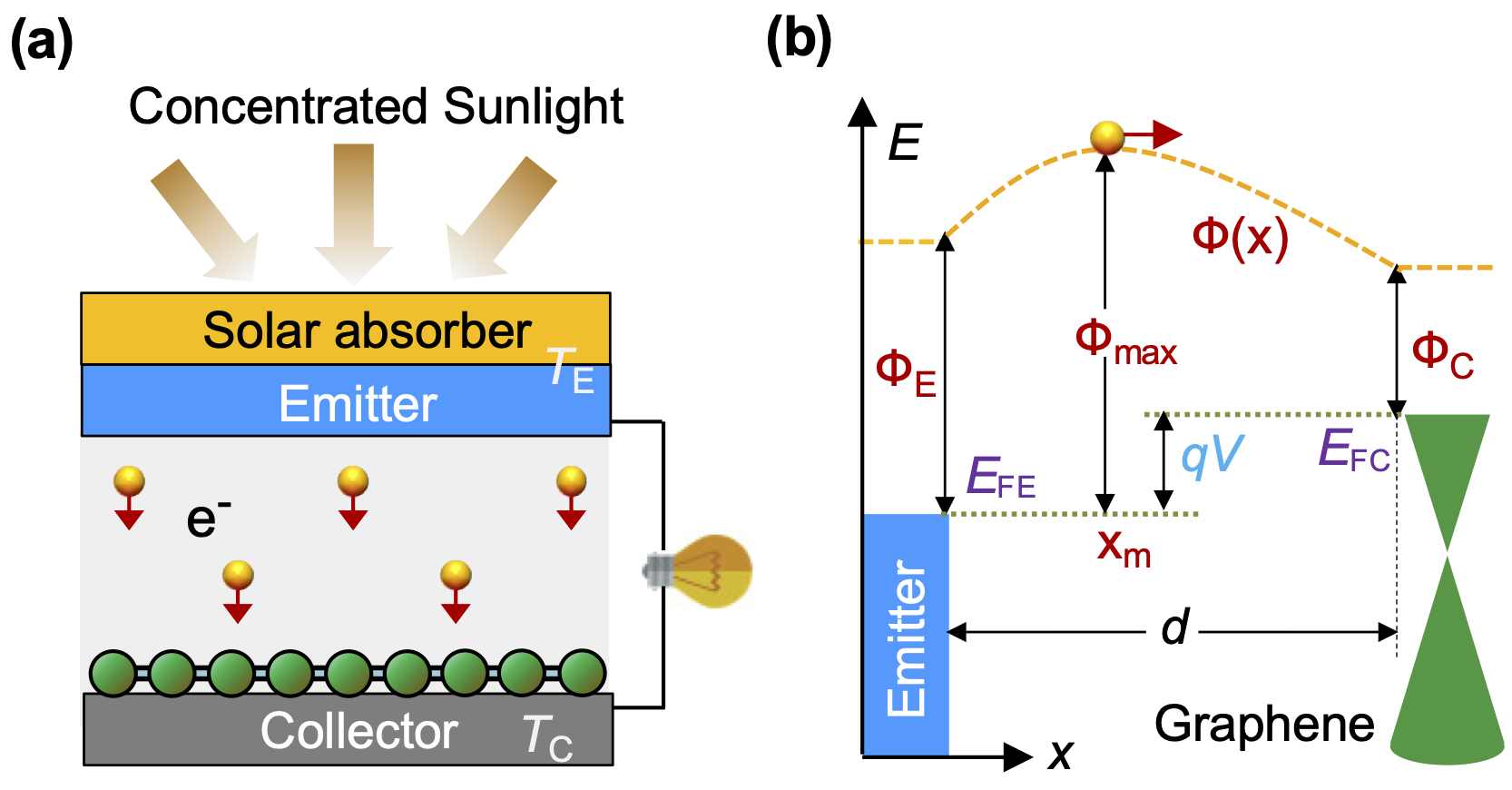}
\caption{(a) Schematic diagram of the thermionic solar cell enabled by back-gated graphene collector. The concentrating solar radiation impinges on the surfaced-textured receiver, feeding the TIEC emitter with thermal energy. The electrons emitted by the emitter due to thermionic emission travel across the vacuum gap, and then condensate at the graphene-based collector. (b) The corresponding energy-band diagram, where $\Phi_{\text{E}}$ ($\Phi_{\text{C}}$) is the work functions of the emitter material (collector), and $E_{\text{FE}}$ ($E_{\text{FC}}$) is the Fermi level of the emitter (collector). The position-dependent potential distribution $\Phi(x)$ along with the $x$ direction yields the maximum value $\Phi_{\text{max}}$. $V$ is the voltage difference between two electrodes and $q$ is the elementary charge.}
\end{figure}

\subsection{Modeling the space charge effect in the interelectrode vacuum gap}
The space-charge effect in a TIEC arises from the Coulombic repulsion caused by the electrons in-transit in interelectrode region. Figure 1(b) illustrates the energy diagram of a space-charge-limited micron-gap vacuum TIEC. The formation of space charge raises the maximum surface potential barrier $\Phi_{\text{max}}$. The space charge effect thus imposes a stricter requirement for electrons to escape from the emitter via the thermionic emission pathway. 
As a result, the space charge effect significantly reduces the output current density of the device.
We model the space charge effect by assuming that electrons transport across the interelectrode vacuum gap is collisionless. 
Under this assumption, the additional barrier height generated by the space charge effect, i.e. $\Phi(x)$, can be obtained from the Poisson's equation
\begin{equation}
\frac{d^2 \Phi}{d x^2}=-q^2 \frac{N(x)}{\varepsilon_{0}},
\end{equation}
where $N(x)=\int_{-\infty}^{+\infty}\,dv_z\int_{-\infty}^{+\infty}\,dv_y\int_{-\infty}^{+\infty}f(x,v)\,dv_x$ denotes the number density of electrons at position $x$, $f(x,v)$ represents the distribution function of electron velocity, and $\varepsilon_{0}$ is the permittivity of free space. 
Assuming that the electron velocity distribution is a half Maxwellian at the position of maximum motive, $x_{\text{m}}$, Eq. (1) can be rewritten in terms of the dimensionless potential barrier, $\gamma$, and the dimensionless distance, $\xi$, as
\begin{equation}
2\frac{d^2 \gamma}{d \xi^2}=\exp(\gamma)[1\pm\text{erf}(\sqrt{\gamma)}],
\end{equation}
where the upper signs apply for $\xi<0$ while lower signs apply for $\xi\geq0$. $\gamma=[\Phi_{\text{max}}-\Phi(x)]/k_{\text{B}}T_{\text{E}}$, $\xi=(x-x_{\text{m}})/x_{\text{0}}$, $x_{\text{0}}=\sqrt{\varepsilon_{0}k_{\text{B}}T_{\text{E}}/2q^2 N(x_{\text{m}})}$, and $\text{erf}(x)=2/\sqrt{\pi}\int_{0}^{x} e^{-t^2}\, dt$ is the error function. $k_{\text{B}}$ is the Boltzmann constant, and $T_{\text{E}}$ is the temperature of the emitter.
Integrating Eq. (2) with the appropriate boundary conditions leads to
\begin{equation}\label{01}
\xi=\mp\int_{0}^{\gamma}\frac{dt}{\sqrt{e^t\pm e^t\text{erf}(\sqrt{t})\mp2\sqrt{t/\pi}-1}}.
\end{equation}
We calculated the value of this integral numerically for a wide range of $\gamma$. The value of $\Phi_{\text{max}}$ depends on the operating voltage, which is listed, for the saturation, space charge, and the retarding modes of operation, as
\begin{equation}
\Phi_{\text{max}}=
\left\{
\begin{array}{lr}
            \Phi_{\text{E}}, &0<V<V_{\text{sat}}  \\
             \Phi_{\text{E}}+\gamma_{\text{E}}k_{\text{B}}T_{\text{E}},  &V_{\text{sat}}<V<V_{\text{cr}} \\
             \Phi_{\text{C}}+qV, &V>V_{\text{cr}} \\
             \end{array}
\right.
\end{equation}
where $\Phi_{\text{E}}$ ($\Phi_{\text{C}}$) is the work function of the emitter (collector), and the output voltage is then given by $qV=E_{\text{FC}}-E_{\text{FE}}$. $\gamma_{\text{E}}$ is the value of $\gamma$ at the emitter surface and is given by $\gamma_{\text{E}}=\ln{(J_{\text{sat}}/J_{\text{E}})}$, in which $J_{\text{sat}}=AT_{\text{E}}^2\exp{[-\Phi_{\text{E}}/(k_{\text{B}}T_{\text{E}})]}$ is the emitter saturation current density and $J_{\text{E}}$ is the emitter current density at voltage $V$. $V_{\text{sat}}$ and $V_{\text{cr}}$ are the saturation and critical-point voltage, respectively \cite{smith2009theory,su2014space}.
Finally, for conventional metallic emitters, the thermionic electrical current density is described by the Richardson-Dushman equation:
\begin{equation}\label{01}
J_{\text{E}}=AT_{\text{E}}^2\exp
\left(
-\frac{\Phi_{\text{max}}}{k_{\text{B}}T_{\text{E}}}
\right),
\end{equation}
where $A$ is a materials-specific Richardson-Dushman constant. 

\subsection{Modelling thermionic emission from graphene: Non-Richardson-Dushman thermionic emission}

In 2D material such as graphene, the thermionic emission physics is radically different from the 3D material counterpart \cite{liang2015electron,rodriguez2016enhanced,misra2017thermionic,sinha2014ideal,ang2018universal,trushin2018theory1,trushin2018theory2}. Firstly, due to the carrier scattering effects, electron momentum conservation is violated during the out-of-plane thermionic emission process \cite{liang2015electron,rodriguez2016enhanced,misra2017thermionic}. Secondly, electrons undergoing thermionic emission are no longer described by the parabolic energy dispersion as in the case of most 3D materials, and is instead described by a linear energy band structure, commonly known as the Dirac cone approximation \cite{sinha2014ideal,ang2018universal,trushin2018theory1,trushin2018theory2}, for carrier energy lower than 1 eV. For electron thermionic emission from graphene that involves higher electron energy as in the case of graphene/vacuum interfacial thermionic emission, the full-band tight-binding energy dispersion should be employed \cite{neto2009electronic,ang2019generalized}. The thermionic current density flowing out of the graphene surface can be modeled as  
\begin{equation}
J=\frac{4q}{(2\pi)^2}\sum_{n} {\frac{v_x^{(n)}(k_x^{(n)})}{t_{\text{Gr}}}}\int\tau^n(\bm{k}_{\lVert},k_x)f_{\text{FD}}(E_{\bm{k}_{\lVert}})\, d^2\bm{k}_{\lVert},
\end{equation}
where the factor $4$ represents the spin-valley degeneracy, $t_{\text{Gr}}=0.335$ nm is the thickness for graphene, $f_{\text{FD}}(E_{\bm{k}_{\lVert}})$ is the Fermi-Dirac distribution function, $\bm{k}_{\lVert}=(k_y,k_z)$ is the electron wave-vector component lying in the 2D $yz$-plane, $x$ denotes the direction orthogonal to the $y-z$ plane of the 2D system, $k_x^{(n)}$ is
the quantized out-of-plane wave-vector component of the $n$-th subband, $v_x^{(n)}(k_x^{(n)})=\sqrt{2mE_{k_x}^{(n)}}$ is the cross-plane electron group velocity, $m$ is the free electron mass, $E_{k_x}$ is the discrete bound-state energy level, and the summation $\sum_{n}$ runs over all of the $n$-th quantized subbands. The
nonconservation of $\bm{k}_{\lVert}$ during the out-of-plane thermionic emission process leads to the coupling between $\bm{k}_{\lVert}$ and $k_x$. Accordingly, the $n$-th subband transmission probability becomes $\tau^{(n)}(\bm{k}_{\lVert},k_x)$; i.e., the cross-plane electron tunneling is dependent on both $\bm{k}_{\lVert}$ and $k_x$.

For thermionic emission, the minimum energy required to overcome the barrier equals $\Phi_{\text{max}}$ so that the transmission probability $\tau^{(n)}(\bm{k}_{\lVert},k_x)=\lambda{\cal H}(E_{\bm{k}_{\lVert}}+E_{k_x}^{(n)}-\Phi_{\text{max}})$; i.e., $E_{\bm{k}_{\lVert}}$ and $E_{k_x}^{(n)}$ are combined to overcome the interface barrier $\Phi_{\text{max}}$. Here, $\lambda$ characterizes the strength of 2D plane momentum-non-conserving scattering process. The term of ${\cal H}(x)$ denotes the Heaviside step function. According to the electronic properties of graphene, the $\bm{k}_{\lVert}$ integral is transformed rewritten as $d^2\bm{k}_{\lVert}=(2\pi)^2D(E_{\bm{k}_{\lVert}})dE_{\bm{k}_{\lVert}}$, where $D(E_{\bm{k}_{\lVert}})$ is the electronic density of states (DOS). For a graphene-vacuum interface barrier, the Fermi-Dirac distribution function approaches the semiclassical Maxwell-Boltzmann distribution function since the emitted electrons are in the nondegenerate regime. Because there is only one set of subbands involve in the thermionic emission from graphene, the $E_{k_x}$ of this subband is conventionally set to zero. This is very commonly performed in the density functional theory (DFT) simulation literature of graphene and its heterostructure, in which the Dirac point is conventionally set as the zero energy. Equations (6) can be simplified into single-subband thermionic emission electrical current density for graphene as
\begin{equation}
J=4qv_x \int\tau(E_{\bm{k}_{\lVert}})\zeta(E_{\bm{k}_{\lVert}})D(E_{\bm{k}_{\lVert}})\, dE_{\bm{k}_{\lVert}},
\end{equation}
where $v_x=\sqrt{2mE_{k_x}}$ is the cross-plane electron group velocity, $\zeta(x)=\exp[-(x-E_{\text{FC}})/k_{\text{B}}T_{\text{C}}]$. In the process of obtaining Eq. (7), the summation sign disappears because in the case of graphene, there exists only one set of spin and valley degenerate subbands that are responsible for the thermionic emission of electrons. Because of the atomic thickness and the semimetallic electronic properties of graphene, there is only one Dirac conic dispersion in the low energy regime around the Fermi level, which evolve into a more complex (and highly nonlinear) band structure in the higher energy regime. This singular set of energy bands thus allow the summation term to be omitted.
The full-band (FB) tight-binding model of graphene yields
\begin{equation}
D(\mu)=
\left\{
\begin{array}{lr}
            \frac{D_0}{\sqrt{\Gamma(\mu)}}{\cal K}(\frac{4\mu}{\Gamma(\mu)}),&0<\mu<1
            \\
             \frac{D_0}{\sqrt{4\mu}}{\cal K}(\frac{\Gamma(\mu)}{4\mu}),&1<\mu<3 \\
             \end{array}
\right.
\end{equation}
where $\mu=E_{\bm{k}_{\lVert}}/E_{0}$, $D_0=(4/A_{\text{c}}\pi^2)(E_{\bm{k}_{\lVert}}/E_{0}^{2})$, $E_{0}=2.8$ eV, $A_{\text{c}}=0.052$ nm$^2$, $\Gamma(x)=(1+x)^2-(x^2-1)^2/4$ , and ${\cal K}(\mu)=\int_{0}^{1} [(1-x^2)(1-\mu x^2)]^{-0.5}\, dx$ is the complete Elliptic integral of the first kind. By taking these into account, the thermionic emission current density from the graphene collector is no longer described by the classic Richardson-Dushman law in Eq. (5), and is, instead, described by the following relation \cite{ang2019generalized}
\begin{equation}
J_{\text{C}}=\tilde{\lambda}\frac{4q v_x}{\pi^2t_{\text{Gr}}E_{0}^{2}}\vartheta(\Phi_{\text{max}})=\frac{4q}{\pi^2\tau_{\text{inj}}E_{0}^{2}}\vartheta(\Phi_{\text{max}}),
\end{equation}
where $\tau_{\text{jnj}}=t_{\text{Gr}}/(\tilde{\lambda} v_x)$ is a charge injection characteristic time constant whose value is influenced by the quality of the contact.
Importantly, recent experiment of graphene/silicon Schottky contact has demonstrated the insufficiency of the Richardson-Dushman model in describing the emission behavior of graphene, and the non-Richardson-Dushman model is found to provide a more appropriate explanation on the observed current-voltage and current-temperature characteristics \cite{javadi2020kinetics}. 
This experimental work thus provide a strong assurance to the use of 2D thermionic emission model as developed above. 
As shown below, $\tau_{inj}$ plays an important role in determining the energy conversion efficiency of the device.

Finally, $\vartheta(\Phi_{\text{max}})$ can be numerically solved from
\begin{eqnarray}\label{02}
&&\vartheta(\Phi_{\text{max}})={\cal H}(E^{\prime}-\Phi_{\text{max}})\nonumber \\
& &\times\int_{\Phi_{\text{max}}}^{E_{0}}\frac{E_{\bm{k}_{\lVert}} dE_{\bm{k}_{\lVert}}}{\sqrt{\Gamma(\mu)}}\zeta(E_{\bm{k}_{\lVert}}){\cal K}\left(\frac{4\mu}{\Gamma(\mu)}\right)\nonumber \\
& &+\int_{E_{0}}^{3E_{0}}\frac{E_{\bm{k}_{\lVert}} dE_{\bm{k}_{\lVert}}}{\sqrt{4\mu}}\zeta(E_{\bm{k}_{\lVert}}){\cal K}\left(\frac{\Gamma(\mu)}{4\mu}\right),
\end{eqnarray}
The net current density, $J$, is the difference between the thermionic current densities from the emitter and the collector, i.e., $J=J_{\text{E}}-J_{\text{C}}$.

\subsection{Modelling conversion efficiency and energy balance}
\begin{figure}
\centering
\includegraphics[height=4cm]{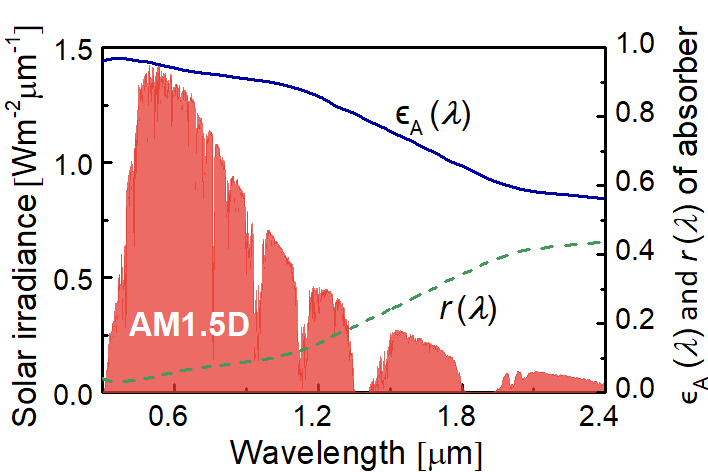}
\caption{Solar irradiance and emissivity of an absorber varying with wavelength.}
\end{figure}
The solar-to-electricity conversion efficiency of a CTSC can be written as the product of the opto-thermal efficiency of the absorber ($\eta_{\text{ot}}$) and the TIEC efficiency ($\eta_{\text{TIEC}}$), which is expressed as $\eta=\eta_{\text{ot}}\eta_{\text{TIEC}}$, where $\eta_{\text{ot}}$ is the efficiency of converting the concentrated solar radiation ($Q_{\text{sun}}$) into heat flux ($Q_{\text{abs}}$) that flows into the emitter, i.e.
\begin{equation}
\eta_{\text{ot}}=\frac{Q_{\text{abs}}}{Q_{\text{sun}}}=\frac{Q_{\text{abs}}}{C\int_{0}^{\infty} I_{\text{AM}}(\lambda)\,d\lambda}.
\end{equation}
Here, $ I_{\text{AM}}(\lambda)$ represents the solar irradiance varying with the wavelength $\lambda$, whose corresponding radiation wavelength of AM 1.5 Direct solar spectra is 0.28 to 2.4 $\mu$m \cite{gueymard2002proposed}, as indicated in Fig. 2. The energy balance equation for the absorber yields $Q_{\text{sun}}=Q_{\text{abs}}+Q_{\text{ref}}+Q_{\text{rad,A}}$, where
\begin{equation}
Q_{\text{ref}}=C\int_{0}^{\infty}r(\lambda) I_{\text{AM}}(\lambda)\, d\lambda
\end{equation}
represents the reflected radiation flow in the absorber, and
\begin{equation}
Q_{\text{rad,A}}=\frac{2\pi h c^2}{\lambda^5} \int_{0}^{\infty}[\Theta(T_{\text{E}})-\Theta(T_{\text{amb}})] I_{\text{AM}}(\lambda)\epsilon_{\text{A}}(\lambda)\, d\lambda
\end{equation}
denotes the spectral blackbody emissive power density into the environment. $\epsilon_{\text{A}}(\lambda)$ is the spectral emissivity of the absorber surface, according to Kirchhoff law, which equals the spectral absorptance, and $r(\lambda)$ stands for the reflection coefficient. Moreover, $C$ is the solar concentration factor, $\Theta(T)=(e^{(\hbar c/\lambda k_{\text{B}}T}-1)^{-1}$ is the Bose-Einstein distributions of photons at the equilibrium temperature, and $T_{\text{amb}}$ is the ambient temperature.
For the thermionic part, the TIEC efficiency is given by 
\begin{equation}\label{04}
\eta_{\text{TIEC}}=\frac{P_{\text{out}}}{Q_{\text{abs}}}=\frac{JV}{Q_{\text{abs}}}.
\end{equation}
where $P_{\text{out}}$ denotes the output electric power density. The energy losses existing in the emitter originate from a number of fundamental energy carriers: thermionically emitted electrons, radiative photons, and electron heat conduction in the leads. The net energy carried by emitted electrons from emitter flowing towards the collector is given by
\begin{equation}\label{04}
Q_{\text{ther,E$\to$C}}=\frac{J\Phi_{\text{max}}+2 k_{\text{B}}T_{\text{E}}J_{\text{E}}-3 k_{\text{B}}T_{\text{C}}J_{\text{C}}}{q}
\end{equation}
The factors $3k_{\text{B}}T$ and $2k_{\text{B}}T$ represent the excess energy of thermionically emitted electrons from the graphene and conventional metal, respectively. It should be noted that the average thermal energy per Dirac fermion is 
\begin{equation}\label{04}
\overline{E}=\frac{\hbar v_{\text{F}}\int_{0}^{\infty} \bm{k}_{\lVert}e^{-\hbar v_{\text{F}}\bm{k}_{\lVert}/k_{\text{B}}T}d\bm{k}_{\lVert}}{\int_{0}^{\infty} e^{-\hbar v_{\text{F}}\bm{k}_{\lVert}/k_{\text{B}}T}d\bm{k}_{\lVert}}=k_{\text{B}}T
\end{equation}
for each degree of freedom, which is different from $k_{\text{B}}T/2$ in metal. 
The total thermal energy is thus $3k_{\text{B}}T$ in Dirac systems with three degrees of freedom, while the thermal energy is $2k_{\text{B}}T$ in typically parabolic systems \cite{liang2017thermionic}. 

The radiative heat transfer between the emitter and the collector is described by the familiar Stefan–Boltzmann law at large inter-electrode gaps:
\begin{equation}\label{06}
Q_{\text{rad,E$\to$C}}=\frac{\sigma(T_{\text{E}}^4-T_{\text{C}}^4)}{1/\epsilon_{\text{E}}+1/\epsilon_{\text{C}}-1},
\end{equation} 
where $\epsilon_{\text{E}}$ and $\epsilon_{\text{C}}$ are the emissivity of the emitter and the collector, and $\sigma$ is the Stefan–Boltzmann constant. Analogously, the collector also includes three major energy losses, such as the electron energy flux due to thermionic emission, radiative energy flux, and heat conduction between the collector and ambient. The net energy carried by emitted electrons from collector flowing towards the emitter is given by
\begin{equation}\label{04}
Q_{\text{ther,C$\to$E}}=\frac{3 k_{\text{B}}T_{\text{C}}J_{\text{C}}-2 k_{\text{B}}T_{\text{E}}J_{\text{E}}-J(\Phi_{\text{max}}-qV)}{q}
\end{equation}
According to the Newton heat transfer law, the thermal loss caused by heat convection and conduction from the external surface of the collector towards the environment is expressed as $Q_{\text{con}}=h(T_{\text{C}}-T_{\text{amb}})$, where $h$ is the global heat transfer coefficient. The temperatures of the emitter and collector, i.e., $T_{\text{E}}$ and $T_{\text{C}}$, are determined by the following energy balance conditions:
\begin{subequations}
\label{eq:whole}
\begin{equation}
Q_{\text{sun}}-Q_{\text{ref}}-Q_{\text{rad,A}}=Q_{\text{ther,E$\to$C}}-Q_{\text{rad,E$\to$C}}, \label{subeq:2}
\end{equation}
\begin{equation}
Q_{\text{ther,C$\to$E}}-Q_{\text{con}}-Q_{\text{rad,E$\to$C}}=0.\label{subeq:1}
\end{equation}
\end{subequations}

In Fig. 3, a flow chart summarizing the self-consistent iterative algorithm to implement the model developed above is shown. The self-consistent algorithm is used to calculate the temperature at different parts of the device as well as the different energy exchange channels. 
The algorithm takes material-and device- related parameters as input at the start of the iterative solution process. 
To start the iteration, an initial guess of the temperatures is provided. 
With this initial guess, the algorithm then checks for convergence of the energy balance criterion at different parts of the device and updates the temperatures accordingly until convergence is achieved. 
Here we note that the various energy exchange channels, such as thermionic and radiative heat flux have strong nonlinear dependences on the electrode temperatures. Therefore, adaptively updated coefficients have been used to update the electrode temperatures during the self-consistent cycles.

\begin{figure}
\centering
\includegraphics[width=6cm]{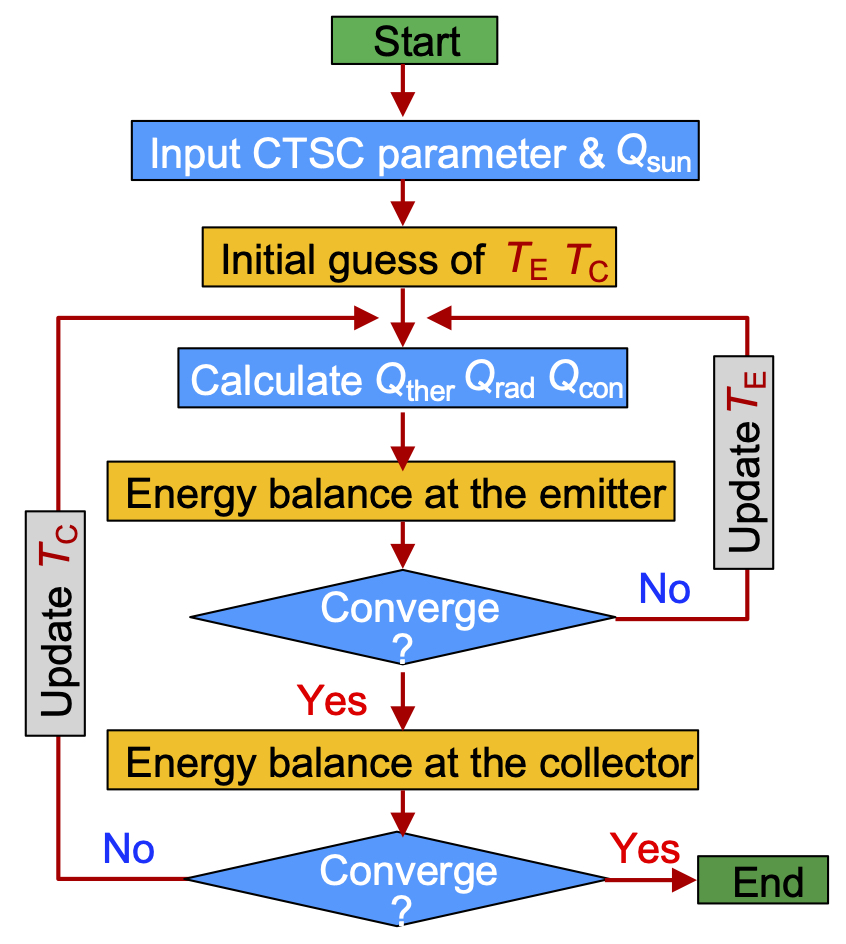}
\caption{Flowchart of the self-consistent numerical iterative model implemented in this work.}
\label{fig:graph}
\end{figure}
\section{Results and Discussion}

Based on the candidate materials of each subcomponents and the 1D energy transfer model described in previous sections, we perform a numerical parameter optimization to obtain a sufficiently realistic estimation of the CSTC system performance and energy conversion efficiency. The work functions of emitter and collector in a TIEC are critically important parameters because electron emission depends on these parameters exponentially. To obtain high thermionic emission at practically achievable temperatures, the emitter work function should be low. 
On the other hand, to maximize efficiency, a large voltage difference between emitter and collector is required, which, in turn, requires the collector work function to be considerably lower than that of the emitter. 
However, if the collector work function is too low, significant back emission from the collector will occur, which lowers the efficiency. 
Given these complex requirements, a reasonable set of values for the emitter and collector work functions is chosen as $\Phi_{\text{E}}=2$ eV and $\Phi_{\text{C}}=1.5$ eV, which forms the base parameters for us to investigate the performance of the thermionic device over a wide range of other parameters. 
Moreover, these values of electrode work functions are also found to be the optimal values for the range of electrode temperatures considered in this study. We assume that: (1) An effective Richardson constant of $A=8.42$ A cm$^{-2}$ K$^{-2}$ is considered in order to estimate the application potential \cite{trucchi2018solar}. (2) The emissivity $\epsilon_{\text{E}}$ of the emitter is assumed to be a function of temperature, which is given by $\epsilon_{\text{E}}=0.13+2.09\times10^{-4}T_{\text{E}}$ \cite{trucchi2018solar}; $\epsilon_{\text{C}}=2.3\%$, namely, the values for the collector (graphene) at the temperature of around 500 K \cite{freitag2010thermal}. (3) The heat transfer coefficient $h$ is assumed to be 0.1 W cm$^{-2}$ K$^{-1}$ \cite{rahman2020interplay}, which is the upper bound of heat transfer coefficient for cooling by free convection.

The performance of a CTSC is strongly influenced by the output voltage. The functional dependence of the operating temperatures, the current density, the energy fluxes and the energy conversion efficiency, thus offer a useful perspective on the operation and performance limit of a CTSC. Based on the energy balance at the emitter and collector, as defined in Eqs. 19(a) and 19(b), respectively, the emitter and collector temperature [Fig. 4(a)], net current density and  maximum  motive [Fig. 4(b)], different energy fluxes [Fig. 4(c)], and efficiency of different components [Fig. 4(d)] are numerically obtained using the self-consistent approach [shown in Fig. 3] for different operating voltages. In the following discussion, the incident spectrum is the AM1.5 direct circumsolar spectrum multiplied by the flux concentration $C=800$. An inter-electrode distance of 2 $\mu$m and $\tau_{\text{inj}}=20$ ns are chosen. The values of these parameters are used unless specifically mentioned.
We derive the current–voltage characteristics of the CTSC. In the saturation region ($0<V<V_{\text{sat}}$), the maximum motive is equal to the emitter work function ($\Phi_{\text{E}}$), which is the minimum energy required by the electron to be thermionically emitted from the emitter. 
All emitted electrons can reach the collector, as there is no potential barrier in the inter-electrode space, and thus the current does not change with voltage. 
In contrast, in the space-charge and retarding regions ($V>V_{\text{cr}}$), the maximum motive is higher than the emitter work function and it gradually rises with the output voltage. 
Consequently, only a portion of the emitted electrons with sufficient energy to overcome the additional energy barrier generated by the space charge effect in the inter-electrode gap can reach the collector. 
As a result, both the current density and the energy flux from the emitter decrease as the voltage increases, leading to an increase in emitter's temperature [see Fig. 4(b)]. This increase in temperature of the emitter raises the radiative heat transfer from the emitter to the collector [see Fig. 4(c)]. This loss mechanism begins to dominate the energy flux from the emitter as the device is driven deeper into the retarding region. In addition, since the energy exchange of the collector is closely dependent on the emitted energy fluxes from the emitter, the collector's temperature and the heat flux released to the ambient from the collector ($Q_{\text{con}}$) decrease monotonically with the increase of the output voltage. 
Such a phenomenon is the direct consequence of the energy balance requirement at the collector. 
The conversion efficiency of the CTSC depends on the output power produced by the TIEC, which is a product of the output current density and the output voltage. 
Due to the interplay of such two parameters, the CTSC reaches an efficiency of up to 12.8\% for a maximum barrier height of 2.9 eV and an output voltage of 1.76 V, which corresponds to the operating temperatures for the emitter and the collector of 1707 K [Fig. 4(c)] and 352 K [Fig. 4(d)], respectively.

\begin{figure*}[htbp]
\centering
\includegraphics[width=18cm]{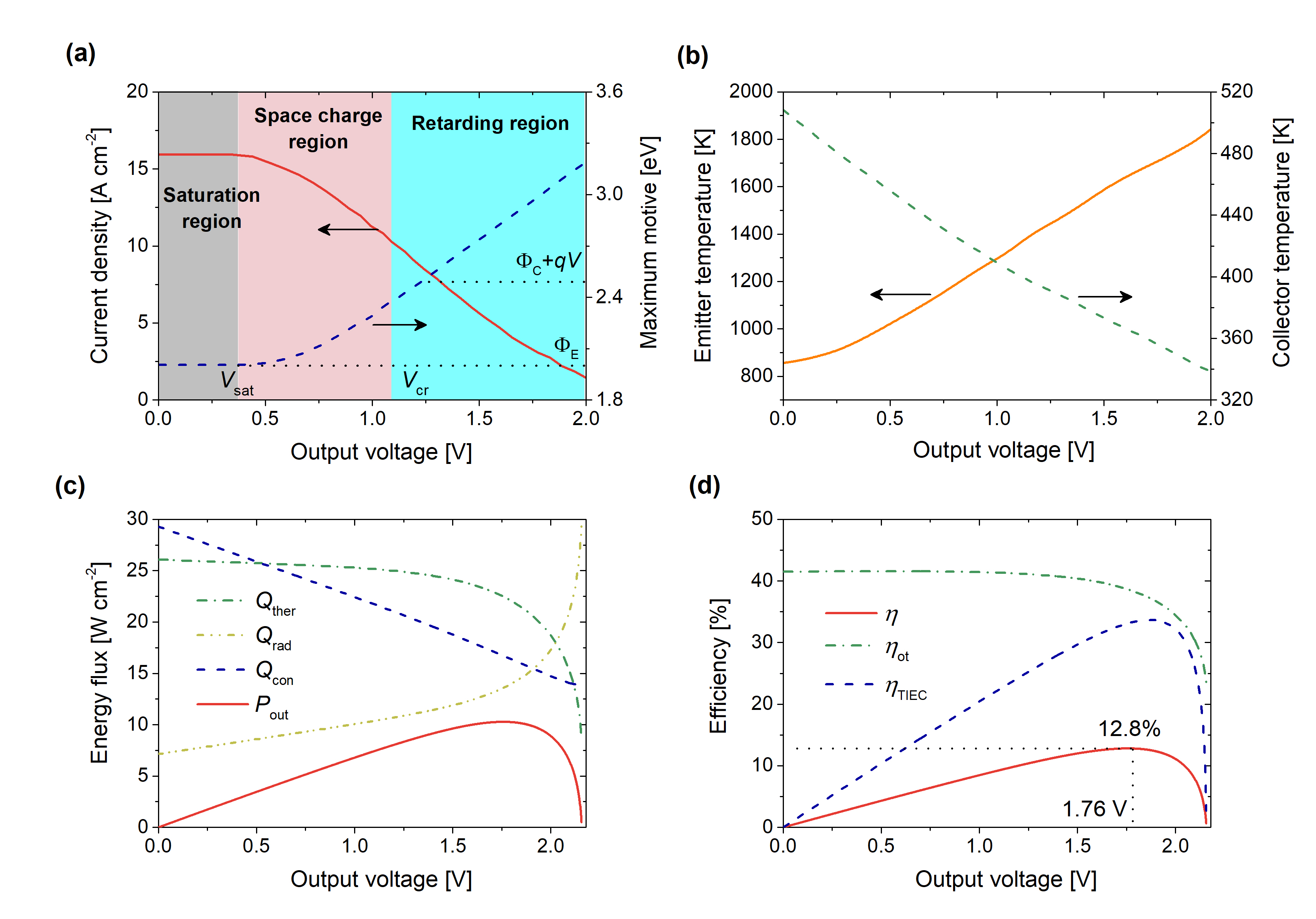}
\caption{(a) current density (solid line) and maximum motive (dash line), (b) electrode temperatures, (c) energy fluxes in the TIEC component, and (d) opto-thermal efficiency of the absorber, TIEC efficiency, and overall conversion efficiency of the CTSC as a function of the operating voltage.}
\label{fig:graph}
\end{figure*}

To provide crystalized guidelines for the performance improvement of practical CTSC system, we study the various processes that underlie the conversion efficiency and the energy loss pathways of the system. Figure 4(d) shows the efficiency of the TIEC and the radiation absorber subsystems as a function of the output voltage. Individually, the opto-thermal efficiency ($\eta_{\text{ot}}$) and the TIEC efficiency ($\eta_{\text{TIEC}}$) reaches 39.8\% and 32.2\%, respectively, thus revealing the TIEC as the more dominant bottleneck that limits the conversion efficiency of the CTSC system. The product of these two individual efficiencies yields the combined energy conversion efficiency of 12.8\% for the CTSC is achievable under 800 sun (1 sun is 0.1 W cm$^{-2}$ K$^{-1}$). In Fig. 4(c), we show the four main energy fluxes existing in the TIEC. We found that the heat conduction and back-body radiative losses are two major loss mechanisms that significantly degrade system performance. To enhance the system performance, more efforts thus should be paid to improve the following components: (1) High-performance wavelength-selective solar receivers with an innovative design uses a high thermal concentration in an evacuated enclosure to prevent air convection and conduction losses. (2) New generation TIEC possesses higher conversion efficiency and operating temperature by fabricating low-work-function thermal stable, and low-emissivity nanostructured materials.

Moreover, we study how different solar concentrations can affect the optimal performance characteristics of the CTSC, as shown in Fig. 5. In general, larger solar concentrations correspond to more input energy fluxes, and thus leading to higher maximum conversion efficiencies ($\eta_{\text{max}}$) [Fig. 5(a)] and electrode temperatures [Fig. 5(b)]. Further increase in efficiency is possible with higher optical concentration for devices to operate at higher temperatures. The optimum values of the maximum motive ($\Phi_{\text{max,opt}}$) and output voltage ($V_{\text{opt}}$) that produces the maximum efficiency also monotonically increase with the solar concentration [Fig. 5 (c) and Fig. 5 (d)]. However, the optimal current density ($J_{\text{opt}}$) decreases with increasing solar concentration. The current density is more strongly influenced by the maximum motive than the temperature (or solar concentration). Specially, when the temperature is relatively low, the effects of maximum motive on the electric current become dominant. Thus, there is an inevitable compromise between the higher conversion efficiency and the higher output power, which may affect the practical design of CTSCs. For example, for high-power energy generator where higher current density and output voltage are desirable, the conversion efficiency could be inherently lower. 

\begin{figure*}[htbp]
\centering
\includegraphics[width=17.2cm]{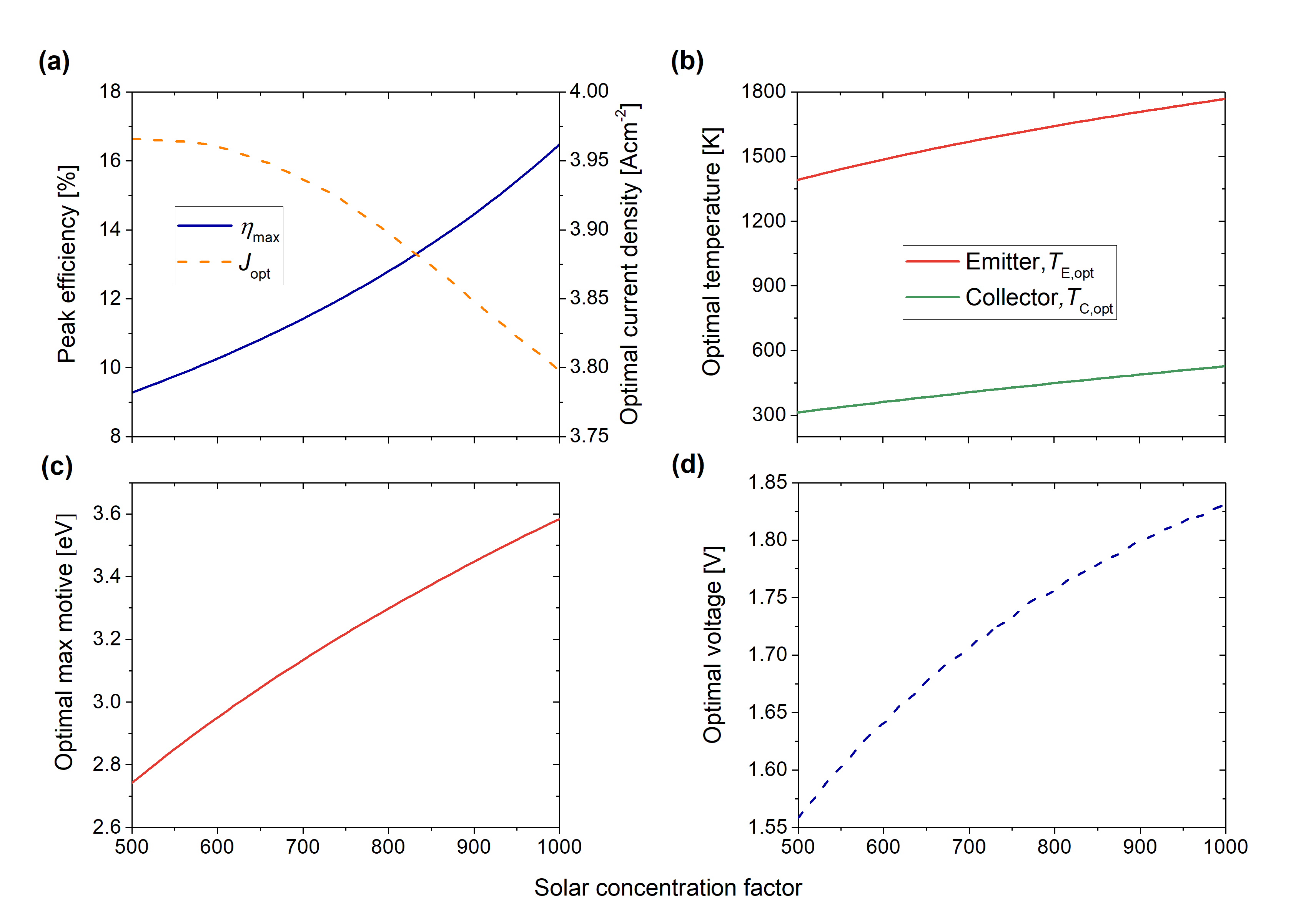}
\caption{(a) The peak efficiency $\eta_{\text{max}}$ (that is, the highest value of the efficiency as a function of voltage) and optimal current density $J_{\text{opt}}$, (b) optimal electrode temperature $T_{\text{E(C),opt}}$, (c) optimal maximum motive $\Phi_{\text{max,opt}}$, and (d) optimal output voltage $V_{\text{opt}}$ under the 500-1000 solar concentrations.}
\label{fig:graph}
\end{figure*}

To investigate the effects of inter-electrode gap width on CTSC operation, the peak efficiency at the inter-electrode spacing region of 2 $\mu$m to 10 $\mu$m are shown in Figs. 6(a), where the output voltages are optimized. It can be seen that the inter-electrode distance has a significant impact on the performance of the CTSC and the peak efficiency can be enhanced when the vacuum gap is decreased. 
As the interelectrode distance increases, the space charge effect starts to dominate, which significantly reduces electron flux from the emitter to the collector. As a result, the output power density decrease at large inte-electrode distances and radiative heat transfer (which is governed by the Stefan-Boltzmann law) gradually becomes dominant due to the rising temperature difference between the electrodes. 
It should be noted that when the gap is very small, the near-field radiative heat transfer between the electrodes, which is caused by the coupling of evanescent waves between the two electrodes, becomes dominant. The energy flux carried by the thermionic electrons is thus small at a very small electrode gap. 
In this case, the current density is low despite mitigation of the space-charge effect. 
We further study the $\tau_{\text{inj}}$-dependence of the peak efficiency in Figs. 6(b). The $\tau_{\text{inj}}$ is related to the out-of-plane velocity of electrons in graphene with nonconserving scattering strength. Here we consider a representative range from 1 ps to 1 ms, which is referenced to the values of graphene-semiconductor contact reported experimentally \cite{sinha2014ideal,massicotte2016photo,trushin2018theory2,javadi2020kinetics}. 
A larger $\tau_{\text{inj}}$ corresponds to the situation in which the contact resistance across the graphene-vacuum interface is large \cite{sinha2014ideal}. In general, the key performance parameters of the CTSC is influenced by the values of $\tau_{\text{inj}}$. 
Particularly when $\tau_{\text{inj}}$ is small, increasing $\tau_{\text{inj}}$ leads to a significant improvement of the peak efficiency [see Fig. 5(a-c)]. Such improvement eventually saturates as $\tau_{\text{inj}}$ is further increased. This analysis thus suggests that electrical contact engineering may offer a route to improve the system performance of CTSCs.

\begin{figure*}[htbp]
\centering
\includegraphics[width=17cm]{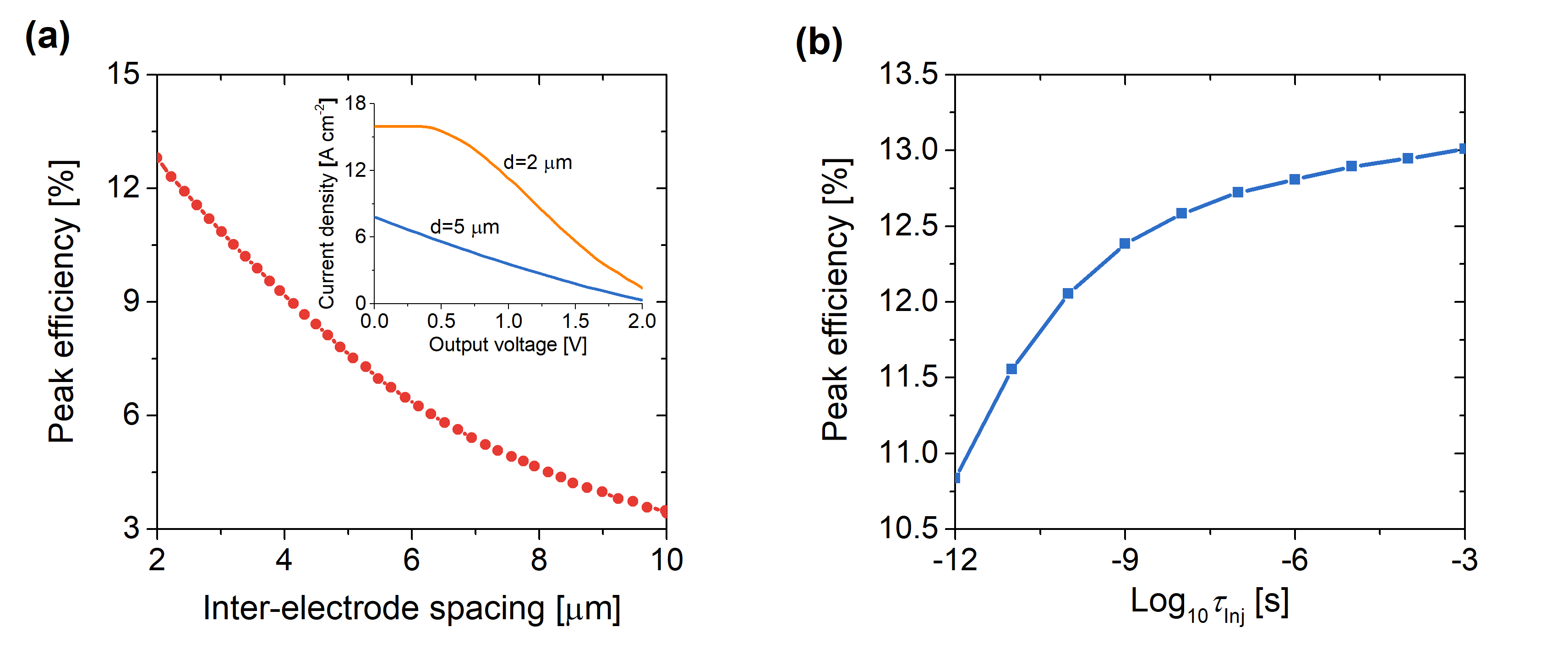}
\caption{(a) The peak efficiency $\eta_{\text{max}}$ varying with inter-electrode spacing $d$ (inset shows a magnified view of the current-voltage graph for $d=2$ $\mu$m and $d=5$ $\mu$m) and charge injection characteristic time $\tau_{\text{inj}}$ under 800 suns. }
\label{fig:graph}
\end{figure*}

To understand the optimal electrode configurations of the thermionic emitter unit, we examine the optimal performances for CTSCs operating in various emitter-collector configurations. Figure 7 shows the peak efficiency for the four different configurations, which are denoted as D-Gr (i.e. diamond films as emitter and graphene as collector), Gr-D (i.e. graphene as emitter and diamond films as collector), Gr-Gr (i.e. graphene as both emitter and collector), and D-D (i.e. diamond films as both emitter and collector). For a solar concentration ranging between 500 to 1000 suns, we found that the best performance is delivered by the D-Gr setup [Fig. 7(a)]. This demonstrates the important role of graphene as a collector material for achieving high-efficiency CTSCs. The physical mechanisms that underlies the better performance of metal-graphene configurations revealed in Fig. 7(a) can be understood as followed. In contrast to diamond- or metal-based collector, graphene quenches the parasitic electron back-flux originating from the collector because of its low electron emission current density – a direct consequence of the vanishing density of states near the Dirac point. Furthermore, low emissivity of graphene avoids heat radiation loss (2.3\%). The combination of these two key factors leads to a much-improved conversion efficiency when graphene is used as a collector material. Fig. 7(b-c) also reveals that: (i) graphene-based emitter generally delivers lowest efficiency because the thermionic emission current is inherently low in graphene; (ii) while thermionic emission current from diamond is substantially larger than that of graphene, the thermionic current backflow and radiative heat loss are also more severe in the D-D configuration. The better conversion efficiency of metal-graphene configuration predicted by our model is also in agreement with a recent experiment in which the overall efficiency using the back-gated graphene collector is found to be 6.7 times higher than that of a TIEC with a tungsten collector \cite{yuan2017back}.

\begin{figure*}[htbp]
\centering
\subfigure{
\includegraphics[width=8cm]{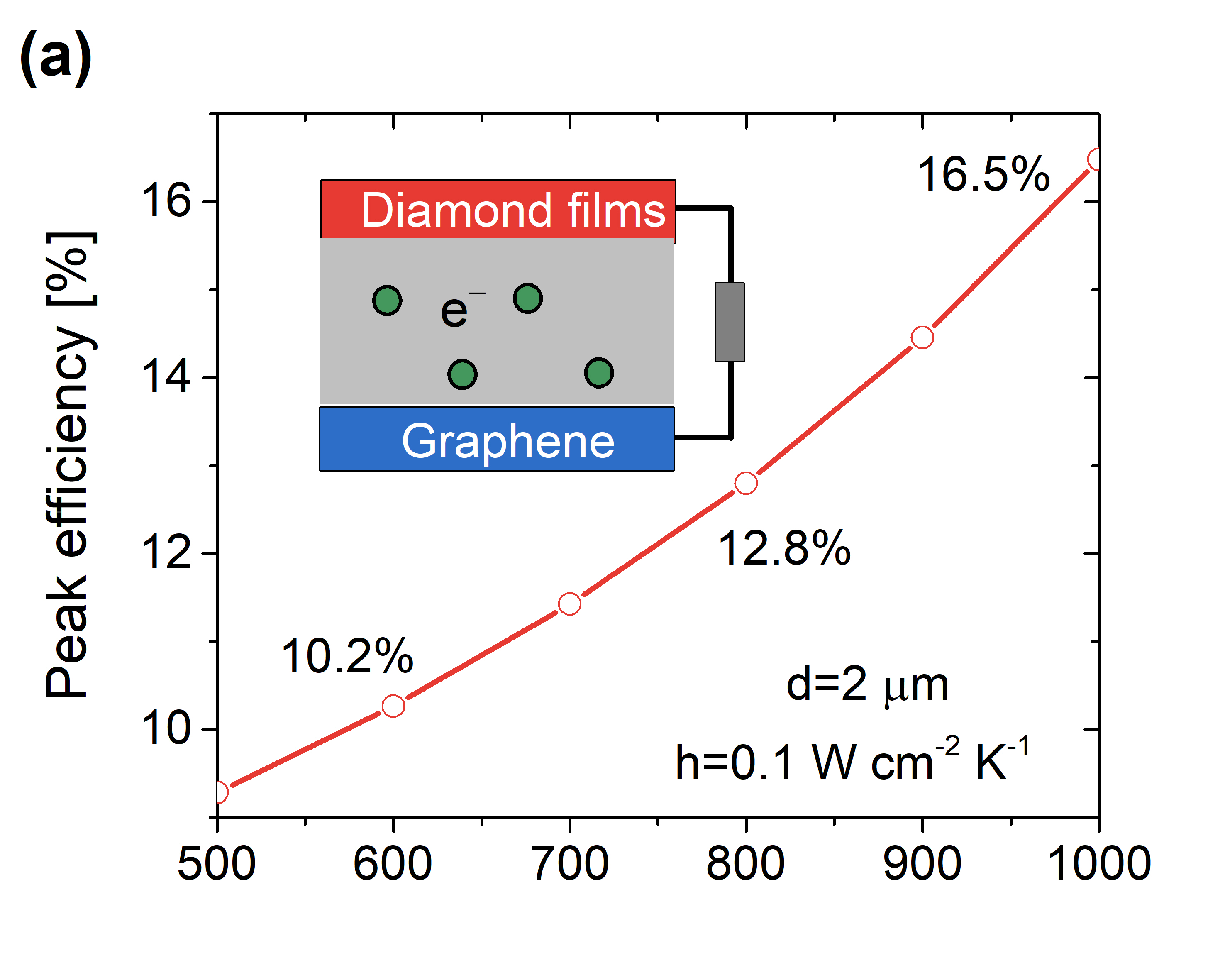}
}
\quad
\subfigure{
\includegraphics[width=8cm]{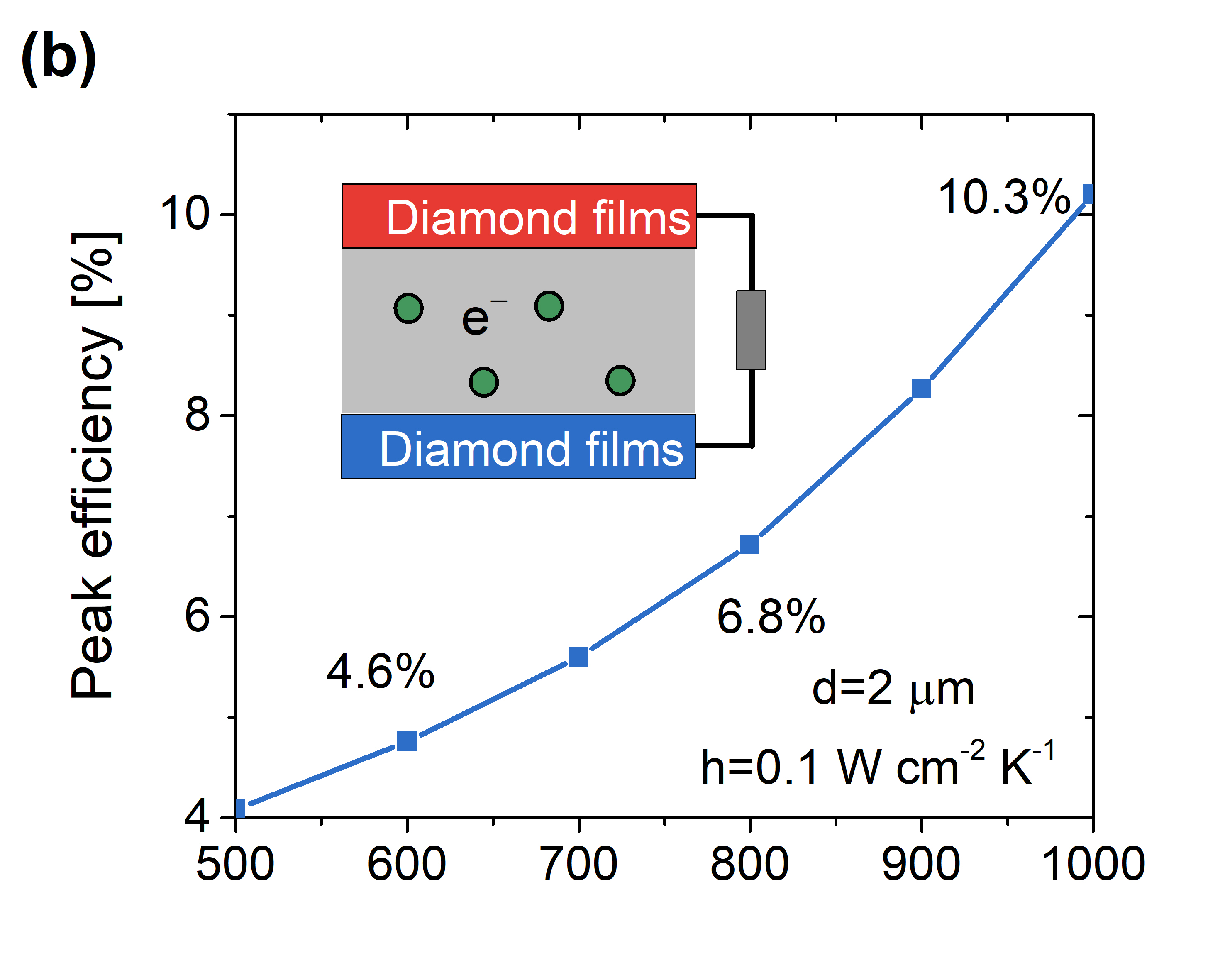}
}
\quad
\subfigure{
\includegraphics[width=8cm]{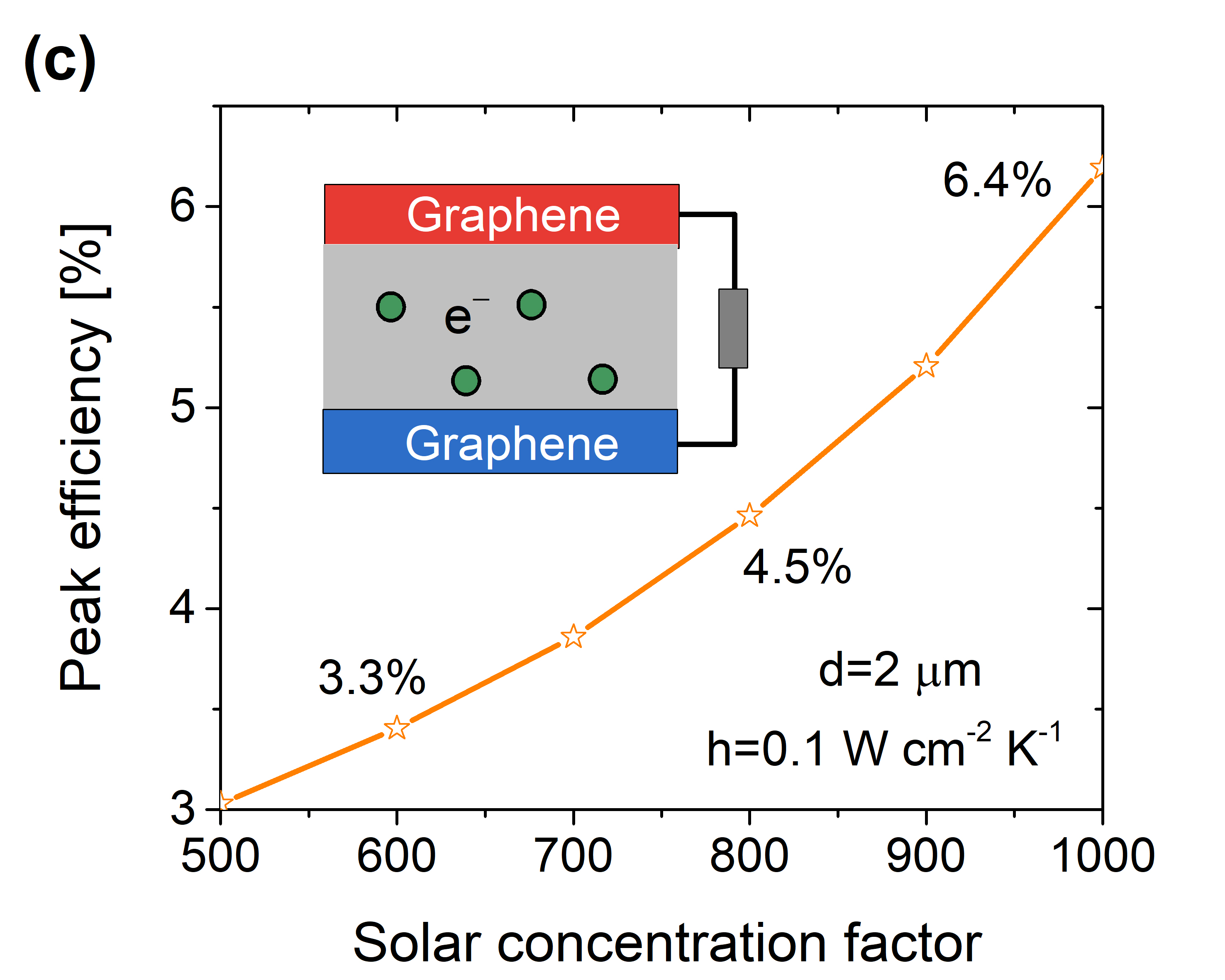}
}
\quad
\subfigure{
\includegraphics[width=8cm]{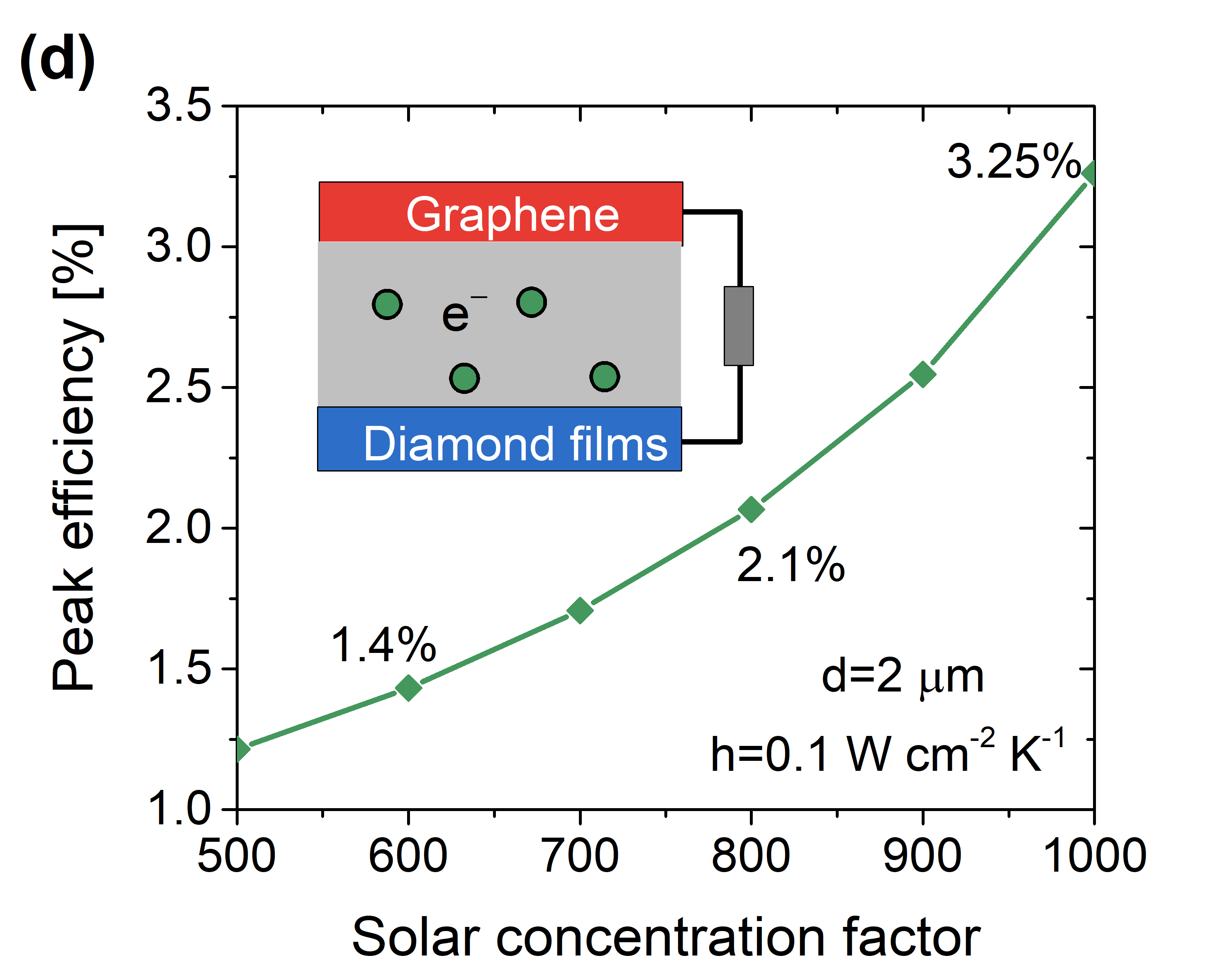}
}
\caption{The peak efficiencies of the four CTSC devices under the 500-1000 solar concentration factors with different emitter-collecotr configurations under the same working conditions, where the output voltage is optimized for each configuration. The four configurations considered are: (a) diamond-graphene; (b) diamond-diamond; (c) graphene-graphene; and (d) graphene-diamond.}
\label{fig:graph}
\end{figure*}

Finally, we briefly comment the potenrial advantages of the proposed system over other concentrating solar power devices, as shown in Table 1, i.e., TEG- \cite{kraemer2016concentrating}, TPV- \cite{bhatt2020high}, and TIEC-based solar cells \cite{rouklove1966thermionic}, we compare the maximum efficiencies and corresponding receiver temperatures regarding these systems, as depicted in Table 1. The theoretically predicted maximum efficiency of the CTSC is substantially higher than those of the solid-state TEG- and TPV-based solar cells, thus establishing the superior solar-to-electricity energy conversion capability of our proposed system.  It should be noted that a prototype TIEC using a barium-tungsten dispenser emitter and a back-gated graphene collector has been experimentally demonstrated to produce a thermionic current density of 10 mA and a conversion efficiency of 9.8\% at emitter temperature of 1375 K \cite{yuan2017back}. Such an efficiency is calculated based on a model that only considers the input power absorbed by the electrons without considers any energy losses. The performance data of the CTSC outlined in Figs. 4-6 and in Table 1 shall offer an important design guideline for the implementations of high-performance CTSCs with different performance characteristics to meet the varying demands of industrial and domestic energy conversions. Importantly, the predicted peak efficiencies shall serve as a useful benchmark value for improving the experimental designs of CTSCs in future experimental investigations.
\begin{table*}
\caption{\label{tab:table1}Performance comparison of representative TEG- \cite{kraemer2016concentrating}, TPV- \cite{bhatt2020high} and TIEC-based \cite{rouklove1966thermionic} solar cells, and our system.}
\begin{ruledtabular}
\begin{tabular}{ccccc}
 System&TEG-based \cite{kraemer2016concentrating}&TPV-based \cite{bhatt2020high}&TIEC-based \cite{rouklove1966thermionic}&CTSC (this work) \\
\hline 
Peak efficiency&7.4\%&8.4\%&7\%&13\% \\
Optimal absorber temperature&975 K&1676 K&2000 K&1687 K \\
\end{tabular}
\end{ruledtabular}
\end{table*}

\section{Conclusion}
In summary, we have established a new concept of the concentrated thermionic solar cells enabled by graphene collector for harvesting solar energy. We developed an analytical model that combines the unconventional thermionic emission characteristics of 2D graphene and the space charge effect in the vacuum gap. It has been found that CTSCs possess a peak efficiency of 12.8\% under 800 sun, which is significantly higher than several existing solar cell architectures, such as thermoelectric and thermophotovoltaic based solar cells, and the optimal conversion efficiency and output electrical power can be customarily made by changing the solar concentrations. By comparing the conversion efficiency of various combinations of graphene and metal in the emitter-collector configurations of the thermionic emission unit, we show that graphene as a collector material delivers the highest conversion efficiency, thus revealing the role of graphene as a key enabler to achieve high efficiency in CTSCs. These findings provide new insights for the design of high-performance CTSCs and shall form the harbinger of 2D-material-based TIEC systems towards cleaner and more sustainable energy.

\section*{Acknowledgment}
This work is supported by the National Natural Science Foundation of China (11675132). LKA and YSA acknowledge the supports of Singapore MOE Tier 2 Grant (2018-T2-1-007).

\bibliography{reference.bib}

\begin{thebibliography}{49}%
\makeatletter
\providecommand \@ifxundefined [1]{%
 \@ifx{#1\undefined}
}%
\providecommand \@ifnum [1]{%
 \ifnum #1\expandafter \@firstoftwo
 \else \expandafter \@secondoftwo
 \fi
}%
\providecommand \@ifx [1]{%
 \ifx #1\expandafter \@firstoftwo
 \else \expandafter \@secondoftwo
 \fi
}%
\providecommand \natexlab [1]{#1}%
\providecommand \enquote  [1]{``#1''}%
\providecommand \bibnamefont  [1]{#1}%
\providecommand \bibfnamefont [1]{#1}%
\providecommand \citenamefont [1]{#1}%
\providecommand \href@noop [0]{\@secondoftwo}%
\providecommand \href [0]{\begingroup \@sanitize@url \@href}%
\providecommand \@href[1]{\@@startlink{#1}\@@href}%
\providecommand \@@href[1]{\endgroup#1\@@endlink}%
\providecommand \@sanitize@url [0]{\catcode `\\12\catcode `\$12\catcode
  `\&12\catcode `\#12\catcode `\^12\catcode `\_12\catcode `\%12\relax}%
\providecommand \@@startlink[1]{}%
\providecommand \@@endlink[0]{}%
\providecommand \url  [0]{\begingroup\@sanitize@url \@url }%
\providecommand \@url [1]{\endgroup\@href {#1}{\urlprefix }}%
\providecommand \urlprefix  [0]{URL }%
\providecommand \Eprint [0]{\href }%
\providecommand \doibase [0]{https://doi.org/}%
\providecommand \selectlanguage [0]{\@gobble}%
\providecommand \bibinfo  [0]{\@secondoftwo}%
\providecommand \bibfield  [0]{\@secondoftwo}%
\providecommand \translation [1]{[#1]}%
\providecommand \BibitemOpen [0]{}%
\providecommand \bibitemStop [0]{}%
\providecommand \bibitemNoStop [0]{.\EOS\space}%
\providecommand \EOS [0]{\spacefactor3000\relax}%
\providecommand \BibitemShut  [1]{\csname bibitem#1\endcsname}%
\let\auto@bib@innerbib\@empty
\bibitem [{\citenamefont {Green}\ and\ \citenamefont
  {Bremner}(2017)}]{green2017energy}%
  \BibitemOpen
  \bibfield  {author} {\bibinfo {author} {\bibfnamefont {M.~A.}\ \bibnamefont
  {Green}}\ and\ \bibinfo {author} {\bibfnamefont {S.~P.}\ \bibnamefont
  {Bremner}},\ }\bibfield  {title} {\bibinfo {title} {Energy conversion
  approaches and materials for high-efficiency photovoltaics},\ }\href@noop {}
  {\bibfield  {journal} {\bibinfo  {journal} {Nat. Mater.}\ }\textbf {\bibinfo
  {volume} {16}},\ \bibinfo {pages} {23} (\bibinfo {year} {2017})}\BibitemShut
  {NoStop}%
\bibitem [{\citenamefont {Gong}\ \emph {et~al.}(2019)\citenamefont {Gong},
  \citenamefont {Li},\ and\ \citenamefont {Wasielewski}}]{gong2019advances}%
  \BibitemOpen
  \bibfield  {author} {\bibinfo {author} {\bibfnamefont {J.}~\bibnamefont
  {Gong}}, \bibinfo {author} {\bibfnamefont {C.}~\bibnamefont {Li}},\ and\
  \bibinfo {author} {\bibfnamefont {M.~R.}\ \bibnamefont {Wasielewski}},\
  }\bibfield  {title} {\bibinfo {title} {Advances in solar energy conversion},\
  }\href@noop {} {\bibfield  {journal} {\bibinfo  {journal} {Chem. Soc. Rev.}\
  }\textbf {\bibinfo {volume} {48}},\ \bibinfo {pages} {1862} (\bibinfo {year}
  {2019})}\BibitemShut {NoStop}%
\bibitem [{\citenamefont {Kamide}\ \emph {et~al.}(2019)\citenamefont {Kamide},
  \citenamefont {Mochizuki}, \citenamefont {Akiyama},\ and\ \citenamefont
  {Takato}}]{kamide2019heat}%
  \BibitemOpen
  \bibfield  {author} {\bibinfo {author} {\bibfnamefont {K.}~\bibnamefont
  {Kamide}}, \bibinfo {author} {\bibfnamefont {T.}~\bibnamefont {Mochizuki}},
  \bibinfo {author} {\bibfnamefont {H.}~\bibnamefont {Akiyama}},\ and\ \bibinfo
  {author} {\bibfnamefont {H.}~\bibnamefont {Takato}},\ }\bibfield  {title}
  {\bibinfo {title} {Heat-recovery solar cell},\ }\href@noop {} {\bibfield
  {journal} {\bibinfo  {journal} {Phys. Rev. Appl.}\ }\textbf {\bibinfo
  {volume} {12}},\ \bibinfo {pages} {064001} (\bibinfo {year}
  {2019})}\BibitemShut {NoStop}%
\bibitem [{\citenamefont {Rau}\ \emph {et~al.}(2017)\citenamefont {Rau},
  \citenamefont {Blank}, \citenamefont {M{\"u}ller},\ and\ \citenamefont
  {Kirchartz}}]{rau2017efficiency}%
  \BibitemOpen
  \bibfield  {author} {\bibinfo {author} {\bibfnamefont {U.}~\bibnamefont
  {Rau}}, \bibinfo {author} {\bibfnamefont {B.}~\bibnamefont {Blank}}, \bibinfo
  {author} {\bibfnamefont {T.~C.}\ \bibnamefont {M{\"u}ller}},\ and\ \bibinfo
  {author} {\bibfnamefont {T.}~\bibnamefont {Kirchartz}},\ }\bibfield  {title}
  {\bibinfo {title} {Efficiency potential of photovoltaic materials and devices
  unveiled by detailed-balance analysis},\ }\href@noop {} {\bibfield  {journal}
  {\bibinfo  {journal} {Phys. Rev. Appl.}\ }\textbf {\bibinfo {volume} {7}},\
  \bibinfo {pages} {044016} (\bibinfo {year} {2017})}\BibitemShut {NoStop}%
\bibitem [{\citenamefont {Guo}\ \emph {et~al.}(2018)\citenamefont {Guo},
  \citenamefont {Luo}, \citenamefont {He}, \citenamefont {Wei},\ and\
  \citenamefont {Li}}]{guo2018photocorrosion}%
  \BibitemOpen
  \bibfield  {author} {\bibinfo {author} {\bibfnamefont {L.-J.}\ \bibnamefont
  {Guo}}, \bibinfo {author} {\bibfnamefont {J.-W.}\ \bibnamefont {Luo}},
  \bibinfo {author} {\bibfnamefont {T.}~\bibnamefont {He}}, \bibinfo {author}
  {\bibfnamefont {S.-H.}\ \bibnamefont {Wei}},\ and\ \bibinfo {author}
  {\bibfnamefont {S.-S.}\ \bibnamefont {Li}},\ }\bibfield  {title} {\bibinfo
  {title} {Photocorrosion-limited maximum efficiency of solar
  photoelectrochemical water splitting},\ }\href@noop {} {\bibfield  {journal}
  {\bibinfo  {journal} {Phys. Rev. Appl.}\ }\textbf {\bibinfo {volume} {10}},\
  \bibinfo {pages} {064059} (\bibinfo {year} {2018})}\BibitemShut {NoStop}%
\bibitem [{\citenamefont {Chester}\ \emph {et~al.}(2011)\citenamefont
  {Chester}, \citenamefont {Bermel}, \citenamefont {Joannopoulos},
  \citenamefont {Soljacic},\ and\ \citenamefont
  {Celanovic}}]{chester2011design}%
  \BibitemOpen
  \bibfield  {author} {\bibinfo {author} {\bibfnamefont {D.}~\bibnamefont
  {Chester}}, \bibinfo {author} {\bibfnamefont {P.}~\bibnamefont {Bermel}},
  \bibinfo {author} {\bibfnamefont {J.~D.}\ \bibnamefont {Joannopoulos}},
  \bibinfo {author} {\bibfnamefont {M.}~\bibnamefont {Soljacic}},\ and\
  \bibinfo {author} {\bibfnamefont {I.}~\bibnamefont {Celanovic}},\ }\bibfield
  {title} {\bibinfo {title} {Design and global optimization of high-efficiency
  solar thermal systems with tungsten cermets},\ }\href@noop {} {\bibfield
  {journal} {\bibinfo  {journal} {Opt. Express}\ }\textbf {\bibinfo {volume}
  {19}},\ \bibinfo {pages} {A245} (\bibinfo {year} {2011})}\BibitemShut
  {NoStop}%
\bibitem [{\citenamefont {Hussain}\ \emph {et~al.}(2018)\citenamefont
  {Hussain}, \citenamefont {M{\'e}n{\'e}zo},\ and\ \citenamefont
  {Kim}}]{hussain2018advances}%
  \BibitemOpen
  \bibfield  {author} {\bibinfo {author} {\bibfnamefont {M.~I.}\ \bibnamefont
  {Hussain}}, \bibinfo {author} {\bibfnamefont {C.}~\bibnamefont
  {M{\'e}n{\'e}zo}},\ and\ \bibinfo {author} {\bibfnamefont {J.-T.}\
  \bibnamefont {Kim}},\ }\bibfield  {title} {\bibinfo {title} {Advances in
  solar thermal harvesting technology based on surface solar absorption
  collectors: A review},\ }\href@noop {} {\bibfield  {journal} {\bibinfo
  {journal} {Sol. Energy Mater. Sol. Cells}\ }\textbf {\bibinfo {volume}
  {187}},\ \bibinfo {pages} {123} (\bibinfo {year} {2018})}\BibitemShut
  {NoStop}%
\bibitem [{\citenamefont {Kraemer}\ \emph {et~al.}(2011)\citenamefont
  {Kraemer}, \citenamefont {Poudel}, \citenamefont {Feng}, \citenamefont
  {Caylor}, \citenamefont {Yu}, \citenamefont {Yan}, \citenamefont {Ma},
  \citenamefont {Wang}, \citenamefont {Wang}, \citenamefont {Muto} \emph
  {et~al.}}]{kraemer2011high}%
  \BibitemOpen
  \bibfield  {author} {\bibinfo {author} {\bibfnamefont {D.}~\bibnamefont
  {Kraemer}}, \bibinfo {author} {\bibfnamefont {B.}~\bibnamefont {Poudel}},
  \bibinfo {author} {\bibfnamefont {H.-P.}\ \bibnamefont {Feng}}, \bibinfo
  {author} {\bibfnamefont {J.~C.}\ \bibnamefont {Caylor}}, \bibinfo {author}
  {\bibfnamefont {B.}~\bibnamefont {Yu}}, \bibinfo {author} {\bibfnamefont
  {X.}~\bibnamefont {Yan}}, \bibinfo {author} {\bibfnamefont {Y.}~\bibnamefont
  {Ma}}, \bibinfo {author} {\bibfnamefont {X.}~\bibnamefont {Wang}}, \bibinfo
  {author} {\bibfnamefont {D.}~\bibnamefont {Wang}}, \bibinfo {author}
  {\bibfnamefont {A.}~\bibnamefont {Muto}}, \emph {et~al.},\ }\bibfield
  {title} {\bibinfo {title} {High-performance flat-panel solar thermoelectric
  generators with high thermal concentration},\ }\href@noop {} {\bibfield
  {journal} {\bibinfo  {journal} {Nat. Mater.}\ }\textbf {\bibinfo {volume}
  {10}},\ \bibinfo {pages} {532} (\bibinfo {year} {2011})}\BibitemShut
  {NoStop}%
\bibitem [{\citenamefont {Lin}\ \emph {et~al.}(2020)\citenamefont {Lin},
  \citenamefont {Lin}, \citenamefont {Yang},\ and\ \citenamefont
  {Jia}}]{lin2020structured}%
  \BibitemOpen
  \bibfield  {author} {\bibinfo {author} {\bibfnamefont {K.-T.}\ \bibnamefont
  {Lin}}, \bibinfo {author} {\bibfnamefont {H.}~\bibnamefont {Lin}}, \bibinfo
  {author} {\bibfnamefont {T.}~\bibnamefont {Yang}},\ and\ \bibinfo {author}
  {\bibfnamefont {B.}~\bibnamefont {Jia}},\ }\bibfield  {title} {\bibinfo
  {title} {Structured graphene metamaterial selective absorbers for high
  efficiency and omnidirectional solar thermal energy conversion},\ }\href@noop
  {} {\bibfield  {journal} {\bibinfo  {journal} {Nat. Commun.}\ }\textbf
  {\bibinfo {volume} {11}},\ \bibinfo {pages} {1} (\bibinfo {year}
  {2020})}\BibitemShut {NoStop}%
\bibitem [{\citenamefont {Khalid}\ \emph {et~al.}(2016)\citenamefont {Khalid},
  \citenamefont {Leong},\ and\ \citenamefont {Mohamed}}]{khalid2016review}%
  \BibitemOpen
  \bibfield  {author} {\bibinfo {author} {\bibfnamefont {K.~A.~A.}\
  \bibnamefont {Khalid}}, \bibinfo {author} {\bibfnamefont {T.~J.}\
  \bibnamefont {Leong}},\ and\ \bibinfo {author} {\bibfnamefont
  {K.}~\bibnamefont {Mohamed}},\ }\bibfield  {title} {\bibinfo {title} {Review
  on thermionic energy converters},\ }\href@noop {} {\bibfield  {journal}
  {\bibinfo  {journal} {IEEE Trans. Electron Devices}\ }\textbf {\bibinfo
  {volume} {63}},\ \bibinfo {pages} {2231} (\bibinfo {year}
  {2016})}\BibitemShut {NoStop}%
\bibitem [{\citenamefont {Zebarjadi}(2017)}]{zebarjadi2017solid}%
  \BibitemOpen
  \bibfield  {author} {\bibinfo {author} {\bibfnamefont {M.}~\bibnamefont
  {Zebarjadi}},\ }\bibfield  {title} {\bibinfo {title} {Solid-state thermionic
  power generators: An analytical analysis in the nonlinear regime},\
  }\href@noop {} {\bibfield  {journal} {\bibinfo  {journal} {Phys. Rev. Appl.}\
  }\textbf {\bibinfo {volume} {8}},\ \bibinfo {pages} {014008} (\bibinfo {year}
  {2017})}\BibitemShut {NoStop}%
\bibitem [{\citenamefont {Zhang}\ \emph {et~al.}(2017)\citenamefont {Zhang},
  \citenamefont {Pan},\ and\ \citenamefont {Chen}}]{zhang2017parametric}%
  \BibitemOpen
  \bibfield  {author} {\bibinfo {author} {\bibfnamefont {X.}~\bibnamefont
  {Zhang}}, \bibinfo {author} {\bibfnamefont {Y.}~\bibnamefont {Pan}},\ and\
  \bibinfo {author} {\bibfnamefont {J.}~\bibnamefont {Chen}},\ }\bibfield
  {title} {\bibinfo {title} {Parametric optimum design of a graphene-based
  thermionic energy converter},\ }\href@noop {} {\bibfield  {journal} {\bibinfo
   {journal} {IEEE Trans. Electron Devices}\ }\textbf {\bibinfo {volume}
  {64}},\ \bibinfo {pages} {4594} (\bibinfo {year} {2017})}\BibitemShut
  {NoStop}%
\bibitem [{\citenamefont {Kaur}\ and\ \citenamefont
  {Johal}(2019)}]{kaur2019thermoelectric}%
  \BibitemOpen
  \bibfield  {author} {\bibinfo {author} {\bibfnamefont {J.}~\bibnamefont
  {Kaur}}\ and\ \bibinfo {author} {\bibfnamefont {R.~S.}\ \bibnamefont
  {Johal}},\ }\bibfield  {title} {\bibinfo {title} {Thermoelectric generator at
  optimal power with external and internal irreversibilities},\ }\href@noop {}
  {\bibfield  {journal} {\bibinfo  {journal} {J. Appl. Phys.}\ }\textbf
  {\bibinfo {volume} {126}},\ \bibinfo {pages} {125111} (\bibinfo {year}
  {2019})}\BibitemShut {NoStop}%
\bibitem [{\citenamefont {Kraemer}\ \emph {et~al.}(2016)\citenamefont
  {Kraemer}, \citenamefont {Jie}, \citenamefont {McEnaney}, \citenamefont
  {Cao}, \citenamefont {Liu}, \citenamefont {Weinstein}, \citenamefont
  {Loomis}, \citenamefont {Ren},\ and\ \citenamefont
  {Chen}}]{kraemer2016concentrating}%
  \BibitemOpen
  \bibfield  {author} {\bibinfo {author} {\bibfnamefont {D.}~\bibnamefont
  {Kraemer}}, \bibinfo {author} {\bibfnamefont {Q.}~\bibnamefont {Jie}},
  \bibinfo {author} {\bibfnamefont {K.}~\bibnamefont {McEnaney}}, \bibinfo
  {author} {\bibfnamefont {F.}~\bibnamefont {Cao}}, \bibinfo {author}
  {\bibfnamefont {W.}~\bibnamefont {Liu}}, \bibinfo {author} {\bibfnamefont
  {L.~A.}\ \bibnamefont {Weinstein}}, \bibinfo {author} {\bibfnamefont
  {J.}~\bibnamefont {Loomis}}, \bibinfo {author} {\bibfnamefont
  {Z.}~\bibnamefont {Ren}},\ and\ \bibinfo {author} {\bibfnamefont
  {G.}~\bibnamefont {Chen}},\ }\bibfield  {title} {\bibinfo {title}
  {Concentrating solar thermoelectric generators with a peak efficiency of
  7.4\%},\ }\href@noop {} {\bibfield  {journal} {\bibinfo  {journal} {Nat.
  Energy}\ }\textbf {\bibinfo {volume} {1}},\ \bibinfo {pages} {1} (\bibinfo
  {year} {2016})}\BibitemShut {NoStop}%
\bibitem [{\citenamefont {Bermel}\ \emph {et~al.}(2010)\citenamefont {Bermel},
  \citenamefont {Ghebrebrhan}, \citenamefont {Chan}, \citenamefont {Yeng},
  \citenamefont {Araghchini}, \citenamefont {Hamam}, \citenamefont {Marton},
  \citenamefont {Jensen}, \citenamefont {Solja{\v{c}}i{\'c}}, \citenamefont
  {Joannopoulos} \emph {et~al.}}]{bermel2010design}%
  \BibitemOpen
  \bibfield  {author} {\bibinfo {author} {\bibfnamefont {P.}~\bibnamefont
  {Bermel}}, \bibinfo {author} {\bibfnamefont {M.}~\bibnamefont {Ghebrebrhan}},
  \bibinfo {author} {\bibfnamefont {W.}~\bibnamefont {Chan}}, \bibinfo {author}
  {\bibfnamefont {Y.~X.}\ \bibnamefont {Yeng}}, \bibinfo {author}
  {\bibfnamefont {M.}~\bibnamefont {Araghchini}}, \bibinfo {author}
  {\bibfnamefont {R.}~\bibnamefont {Hamam}}, \bibinfo {author} {\bibfnamefont
  {C.~H.}\ \bibnamefont {Marton}}, \bibinfo {author} {\bibfnamefont {K.~F.}\
  \bibnamefont {Jensen}}, \bibinfo {author} {\bibfnamefont {M.}~\bibnamefont
  {Solja{\v{c}}i{\'c}}}, \bibinfo {author} {\bibfnamefont {J.~D.}\ \bibnamefont
  {Joannopoulos}}, \emph {et~al.},\ }\bibfield  {title} {\bibinfo {title}
  {Design and global optimization of high-efficiency thermophotovoltaic
  systems},\ }\href@noop {} {\bibfield  {journal} {\bibinfo  {journal} {Opt.
  Express}\ }\textbf {\bibinfo {volume} {18}},\ \bibinfo {pages} {A314}
  (\bibinfo {year} {2010})}\BibitemShut {NoStop}%
\bibitem [{\citenamefont {Bhatt}\ \emph {et~al.}(2020)\citenamefont {Bhatt},
  \citenamefont {Kravchenko},\ and\ \citenamefont {Gupta}}]{bhatt2020high}%
  \BibitemOpen
  \bibfield  {author} {\bibinfo {author} {\bibfnamefont {R.}~\bibnamefont
  {Bhatt}}, \bibinfo {author} {\bibfnamefont {I.}~\bibnamefont {Kravchenko}},\
  and\ \bibinfo {author} {\bibfnamefont {M.}~\bibnamefont {Gupta}},\ }\bibfield
   {title} {\bibinfo {title} {High-efficiency solar thermophotovoltaic system
  using a nanostructure-based selective emitter},\ }\href@noop {} {\bibfield
  {journal} {\bibinfo  {journal} {Solar Energy}\ }\textbf {\bibinfo {volume}
  {197}},\ \bibinfo {pages} {538} (\bibinfo {year} {2020})}\BibitemShut
  {NoStop}%
\bibitem [{\citenamefont {Svetovoy}\ and\ \citenamefont
  {Palasantzas}(2014)}]{svetovoy2014graphene}%
  \BibitemOpen
  \bibfield  {author} {\bibinfo {author} {\bibfnamefont {V.}~\bibnamefont
  {Svetovoy}}\ and\ \bibinfo {author} {\bibfnamefont {G.}~\bibnamefont
  {Palasantzas}},\ }\bibfield  {title} {\bibinfo {title} {Graphene-on-silicon
  near-field thermophotovoltaic cell},\ }\href@noop {} {\bibfield  {journal}
  {\bibinfo  {journal} {Phys. Rev. Appl.}\ }\textbf {\bibinfo {volume} {2}},\
  \bibinfo {pages} {034006} (\bibinfo {year} {2014})}\BibitemShut {NoStop}%
\bibitem [{\citenamefont {Yuan}\ \emph {et~al.}(2017)\citenamefont {Yuan},
  \citenamefont {Riley}, \citenamefont {Shen}, \citenamefont {Pianetta},
  \citenamefont {Melosh},\ and\ \citenamefont {Howe}}]{yuan2017back}%
  \BibitemOpen
  \bibfield  {author} {\bibinfo {author} {\bibfnamefont {H.}~\bibnamefont
  {Yuan}}, \bibinfo {author} {\bibfnamefont {D.~C.}\ \bibnamefont {Riley}},
  \bibinfo {author} {\bibfnamefont {Z.-X.}\ \bibnamefont {Shen}}, \bibinfo
  {author} {\bibfnamefont {P.~A.}\ \bibnamefont {Pianetta}}, \bibinfo {author}
  {\bibfnamefont {N.~A.}\ \bibnamefont {Melosh}},\ and\ \bibinfo {author}
  {\bibfnamefont {R.~T.}\ \bibnamefont {Howe}},\ }\bibfield  {title} {\bibinfo
  {title} {Back-gated graphene anode for more efficient thermionic energy
  converters},\ }\href@noop {} {\bibfield  {journal} {\bibinfo  {journal} {Nano
  Energy}\ }\textbf {\bibinfo {volume} {32}},\ \bibinfo {pages} {67} (\bibinfo
  {year} {2017})}\BibitemShut {NoStop}%
\bibitem [{\citenamefont {Gubbels}\ \emph {et~al.}(1989)\citenamefont
  {Gubbels}, \citenamefont {Wolff},\ and\ \citenamefont
  {Metselaar}}]{gubbels1989wf6}%
  \BibitemOpen
  \bibfield  {author} {\bibinfo {author} {\bibfnamefont {G.}~\bibnamefont
  {Gubbels}}, \bibinfo {author} {\bibfnamefont {L.}~\bibnamefont {Wolff}},\
  and\ \bibinfo {author} {\bibfnamefont {R.}~\bibnamefont {Metselaar}},\
  }\bibfield  {title} {\bibinfo {title} {A wf6-cvd tungsten film as an emitter
  for a thermionic energy converter: Ii. electron and cs+-ion emission from
  wf6-cvd tungsten films},\ }\href@noop {} {\bibfield  {journal} {\bibinfo
  {journal} {Appl. Surf. Sci.}\ }\textbf {\bibinfo {volume} {40}},\ \bibinfo
  {pages} {201} (\bibinfo {year} {1989})}\BibitemShut {NoStop}%
\bibitem [{\citenamefont {Liang}\ and\ \citenamefont
  {Ang}(2015)}]{liang2015electron}%
  \BibitemOpen
  \bibfield  {author} {\bibinfo {author} {\bibfnamefont {S.-J.}\ \bibnamefont
  {Liang}}\ and\ \bibinfo {author} {\bibfnamefont {L.}~\bibnamefont {Ang}},\
  }\bibfield  {title} {\bibinfo {title} {Electron thermionic emission from
  graphene and a thermionic energy converter},\ }\href@noop {} {\bibfield
  {journal} {\bibinfo  {journal} {Phys. Rev. Appl.}\ }\textbf {\bibinfo
  {volume} {3}},\ \bibinfo {pages} {014002} (\bibinfo {year}
  {2015})}\BibitemShut {NoStop}%
\bibitem [{\citenamefont {Smith}\ \emph {et~al.}(2009)\citenamefont {Smith},
  \citenamefont {Bilbro},\ and\ \citenamefont {Nemanich}}]{smith2009theory}%
  \BibitemOpen
  \bibfield  {author} {\bibinfo {author} {\bibfnamefont {J.~R.}\ \bibnamefont
  {Smith}}, \bibinfo {author} {\bibfnamefont {G.~L.}\ \bibnamefont {Bilbro}},\
  and\ \bibinfo {author} {\bibfnamefont {R.~J.}\ \bibnamefont {Nemanich}},\
  }\bibfield  {title} {\bibinfo {title} {Theory of space charge limited regime
  of thermionic energy converter with negative electron affinity emitter},\
  }\href@noop {} {\bibfield  {journal} {\bibinfo  {journal} {J. Vac. Sci.
  Technol. B}\ }\textbf {\bibinfo {volume} {27}},\ \bibinfo {pages} {1132}
  (\bibinfo {year} {2009})}\BibitemShut {NoStop}%
\bibitem [{\citenamefont {Su}\ \emph {et~al.}(2014)\citenamefont {Su},
  \citenamefont {Wang}, \citenamefont {Liu}, \citenamefont {Su},\ and\
  \citenamefont {Chen}}]{su2014space}%
  \BibitemOpen
  \bibfield  {author} {\bibinfo {author} {\bibfnamefont {S.}~\bibnamefont
  {Su}}, \bibinfo {author} {\bibfnamefont {Y.}~\bibnamefont {Wang}}, \bibinfo
  {author} {\bibfnamefont {T.}~\bibnamefont {Liu}}, \bibinfo {author}
  {\bibfnamefont {G.}~\bibnamefont {Su}},\ and\ \bibinfo {author}
  {\bibfnamefont {J.}~\bibnamefont {Chen}},\ }\bibfield  {title} {\bibinfo
  {title} {Space charge effects on the maximum efficiency and parametric design
  of a photon-enhanced thermionic solar cell},\ }\href@noop {} {\bibfield
  {journal} {\bibinfo  {journal} {Sol. Energy Mater. Sol. Cells}\ }\textbf
  {\bibinfo {volume} {121}},\ \bibinfo {pages} {137} (\bibinfo {year}
  {2014})}\BibitemShut {NoStop}%
\bibitem [{\citenamefont {Yuan}\ \emph {et~al.}(2015)\citenamefont {Yuan},
  \citenamefont {Chang}, \citenamefont {Bargatin}, \citenamefont {Wang},
  \citenamefont {Riley}, \citenamefont {Wang}, \citenamefont {Schwede},
  \citenamefont {Provine}, \citenamefont {Pop}, \citenamefont {Shen} \emph
  {et~al.}}]{yuan2015engineering}%
  \BibitemOpen
  \bibfield  {author} {\bibinfo {author} {\bibfnamefont {H.}~\bibnamefont
  {Yuan}}, \bibinfo {author} {\bibfnamefont {S.}~\bibnamefont {Chang}},
  \bibinfo {author} {\bibfnamefont {I.}~\bibnamefont {Bargatin}}, \bibinfo
  {author} {\bibfnamefont {N.~C.}\ \bibnamefont {Wang}}, \bibinfo {author}
  {\bibfnamefont {D.~C.}\ \bibnamefont {Riley}}, \bibinfo {author}
  {\bibfnamefont {H.}~\bibnamefont {Wang}}, \bibinfo {author} {\bibfnamefont
  {J.~W.}\ \bibnamefont {Schwede}}, \bibinfo {author} {\bibfnamefont
  {J.}~\bibnamefont {Provine}}, \bibinfo {author} {\bibfnamefont
  {E.}~\bibnamefont {Pop}}, \bibinfo {author} {\bibfnamefont {Z.-X.}\
  \bibnamefont {Shen}}, \emph {et~al.},\ }\bibfield  {title} {\bibinfo {title}
  {Engineering ultra-low work function of graphene},\ }\href@noop {} {\bibfield
   {journal} {\bibinfo  {journal} {Nano Lett.}\ }\textbf {\bibinfo {volume}
  {15}},\ \bibinfo {pages} {6475} (\bibinfo {year} {2015})}\BibitemShut
  {NoStop}%
\bibitem [{\citenamefont {Zhang}\ \emph {et~al.}(2018)\citenamefont {Zhang},
  \citenamefont {Zhang}, \citenamefont {Ye}, \citenamefont {Li}, \citenamefont
  {Liao},\ and\ \citenamefont {Chen}}]{zhang2018graphene}%
  \BibitemOpen
  \bibfield  {author} {\bibinfo {author} {\bibfnamefont {X.}~\bibnamefont
  {Zhang}}, \bibinfo {author} {\bibfnamefont {Y.}~\bibnamefont {Zhang}},
  \bibinfo {author} {\bibfnamefont {Z.}~\bibnamefont {Ye}}, \bibinfo {author}
  {\bibfnamefont {W.}~\bibnamefont {Li}}, \bibinfo {author} {\bibfnamefont
  {T.}~\bibnamefont {Liao}},\ and\ \bibinfo {author} {\bibfnamefont
  {J.}~\bibnamefont {Chen}},\ }\bibfield  {title} {\bibinfo {title}
  {Graphene-based thermionic solar cells},\ }\href@noop {} {\bibfield
  {journal} {\bibinfo  {journal} {IEEE Electron Device Lett.}\ }\textbf
  {\bibinfo {volume} {39}},\ \bibinfo {pages} {383} (\bibinfo {year}
  {2018})}\BibitemShut {NoStop}%
\bibitem [{\citenamefont {Sani}\ \emph {et~al.}(2012)\citenamefont {Sani},
  \citenamefont {Mercatelli}, \citenamefont {Sansoni}, \citenamefont
  {Silvestroni},\ and\ \citenamefont {Sciti}}]{sani2012spectrally}%
  \BibitemOpen
  \bibfield  {author} {\bibinfo {author} {\bibfnamefont {E.}~\bibnamefont
  {Sani}}, \bibinfo {author} {\bibfnamefont {L.}~\bibnamefont {Mercatelli}},
  \bibinfo {author} {\bibfnamefont {P.}~\bibnamefont {Sansoni}}, \bibinfo
  {author} {\bibfnamefont {L.}~\bibnamefont {Silvestroni}},\ and\ \bibinfo
  {author} {\bibfnamefont {D.}~\bibnamefont {Sciti}},\ }\bibfield  {title}
  {\bibinfo {title} {Spectrally selective ultra-high temperature ceramic
  absorbers for high-temperature solar plants},\ }\href@noop {} {\bibfield
  {journal} {\bibinfo  {journal} {J. Renewable Sustainable Energ}\ }\textbf
  {\bibinfo {volume} {4}},\ \bibinfo {pages} {033104} (\bibinfo {year}
  {2012})}\BibitemShut {NoStop}%
\bibitem [{\citenamefont {Hans}\ \emph {et~al.}(2018)\citenamefont {Hans},
  \citenamefont {Latha}, \citenamefont {Bera},\ and\ \citenamefont
  {Barshilia}}]{hans2018hafnium}%
  \BibitemOpen
  \bibfield  {author} {\bibinfo {author} {\bibfnamefont {K.}~\bibnamefont
  {Hans}}, \bibinfo {author} {\bibfnamefont {S.}~\bibnamefont {Latha}},
  \bibinfo {author} {\bibfnamefont {P.}~\bibnamefont {Bera}},\ and\ \bibinfo
  {author} {\bibfnamefont {H.~C.}\ \bibnamefont {Barshilia}},\ }\bibfield
  {title} {\bibinfo {title} {Hafnium carbide based solar absorber coatings with
  high spectral selectivity},\ }\href@noop {} {\bibfield  {journal} {\bibinfo
  {journal} {Sol. Energy Mater. Sol. Cells}\ }\textbf {\bibinfo {volume}
  {185}},\ \bibinfo {pages} {1} (\bibinfo {year} {2018})}\BibitemShut {NoStop}%
\bibitem [{\citenamefont {Koeck}\ and\ \citenamefont
  {Nemanich}(2012)}]{koeck2012substrate}%
  \BibitemOpen
  \bibfield  {author} {\bibinfo {author} {\bibfnamefont {F.~A.}\ \bibnamefont
  {Koeck}}\ and\ \bibinfo {author} {\bibfnamefont {R.~J.}\ \bibnamefont
  {Nemanich}},\ }\bibfield  {title} {\bibinfo {title} {Substrate-diamond
  interface considerations for enhanced thermionic electron emission from
  nitrogen doped diamond films},\ }\href@noop {} {\bibfield  {journal}
  {\bibinfo  {journal} {J. Appl. Phys.}\ }\textbf {\bibinfo {volume} {112}},\
  \bibinfo {pages} {113707} (\bibinfo {year} {2012})}\BibitemShut {NoStop}%
\bibitem [{\citenamefont {Koeck}\ \emph {et~al.}(2011)\citenamefont {Koeck},
  \citenamefont {Nemanich}, \citenamefont {Balasubramaniam}, \citenamefont
  {Haenen},\ and\ \citenamefont {Sharp}}]{koeck2011enhanced}%
  \BibitemOpen
  \bibfield  {author} {\bibinfo {author} {\bibfnamefont {F.~A.}\ \bibnamefont
  {Koeck}}, \bibinfo {author} {\bibfnamefont {R.~J.}\ \bibnamefont {Nemanich}},
  \bibinfo {author} {\bibfnamefont {Y.}~\bibnamefont {Balasubramaniam}},
  \bibinfo {author} {\bibfnamefont {K.}~\bibnamefont {Haenen}},\ and\ \bibinfo
  {author} {\bibfnamefont {J.}~\bibnamefont {Sharp}},\ }\bibfield  {title}
  {\bibinfo {title} {Enhanced thermionic energy conversion and thermionic
  emission from doped diamond films through methane exposure},\ }\href@noop {}
  {\bibfield  {journal} {\bibinfo  {journal} {Diamond Relat. Mater.}\ }\textbf
  {\bibinfo {volume} {20}},\ \bibinfo {pages} {1229} (\bibinfo {year}
  {2011})}\BibitemShut {NoStop}%
\bibitem [{\citenamefont {Nemanich}\ \emph {et~al.}(2014)\citenamefont
  {Nemanich}, \citenamefont {Carlisle}, \citenamefont {Hirata},\ and\
  \citenamefont {Haenen}}]{nemanich2014cvd}%
  \BibitemOpen
  \bibfield  {author} {\bibinfo {author} {\bibfnamefont {R.~J.}\ \bibnamefont
  {Nemanich}}, \bibinfo {author} {\bibfnamefont {J.~A.}\ \bibnamefont
  {Carlisle}}, \bibinfo {author} {\bibfnamefont {A.}~\bibnamefont {Hirata}},\
  and\ \bibinfo {author} {\bibfnamefont {K.}~\bibnamefont {Haenen}},\
  }\bibfield  {title} {\bibinfo {title} {Cvd diamond—research, applications,
  and challenges},\ }\href@noop {} {\bibfield  {journal} {\bibinfo  {journal}
  {MRS Bullet.}\ }\textbf {\bibinfo {volume} {39}},\ \bibinfo {pages} {490}
  (\bibinfo {year} {2014})}\BibitemShut {NoStop}%
\bibitem [{\citenamefont {Novoselov}\ \emph {et~al.}(2005)\citenamefont
  {Novoselov}, \citenamefont {Geim}, \citenamefont {Morozov}, \citenamefont
  {Jiang}, \citenamefont {Katsnelson}, \citenamefont {Grigorieva},
  \citenamefont {Dubonos},\ and\ \citenamefont {Firsov}}]{novoselov2005two}%
  \BibitemOpen
  \bibfield  {author} {\bibinfo {author} {\bibfnamefont {K.~S.}\ \bibnamefont
  {Novoselov}}, \bibinfo {author} {\bibfnamefont {A.~K.}\ \bibnamefont {Geim}},
  \bibinfo {author} {\bibfnamefont {S.~V.}\ \bibnamefont {Morozov}}, \bibinfo
  {author} {\bibfnamefont {D.}~\bibnamefont {Jiang}}, \bibinfo {author}
  {\bibfnamefont {M.~I.}\ \bibnamefont {Katsnelson}}, \bibinfo {author}
  {\bibfnamefont {I.}~\bibnamefont {Grigorieva}}, \bibinfo {author}
  {\bibfnamefont {S.}~\bibnamefont {Dubonos}},\ and\ \bibinfo {author}
  {\bibfnamefont {A.}~\bibnamefont {Firsov}},\ }\bibfield  {title} {\bibinfo
  {title} {Two-dimensional gas of massless dirac fermions in graphene},\
  }\href@noop {} {\bibfield  {journal} {\bibinfo  {journal} {Nature}\ }\textbf
  {\bibinfo {volume} {438}},\ \bibinfo {pages} {197} (\bibinfo {year}
  {2005})}\BibitemShut {NoStop}%
\bibitem [{\citenamefont {Mortazavi}(2017)}]{mortazavi2017ultra}%
  \BibitemOpen
  \bibfield  {author} {\bibinfo {author} {\bibfnamefont {B.}~\bibnamefont
  {Mortazavi}},\ }\bibfield  {title} {\bibinfo {title} {Ultra high stiffness
  and thermal conductivity of graphene like c3n},\ }\href@noop {} {\bibfield
  {journal} {\bibinfo  {journal} {Carbon}\ }\textbf {\bibinfo {volume} {118}},\
  \bibinfo {pages} {25} (\bibinfo {year} {2017})}\BibitemShut {NoStop}%
\bibitem [{\citenamefont {Zamani}\ and\ \citenamefont
  {Farghadan}(2018)}]{zamani2018graphene}%
  \BibitemOpen
  \bibfield  {author} {\bibinfo {author} {\bibfnamefont {S.}~\bibnamefont
  {Zamani}}\ and\ \bibinfo {author} {\bibfnamefont {R.}~\bibnamefont
  {Farghadan}},\ }\bibfield  {title} {\bibinfo {title} {Graphene nanoribbon
  spin-photodetector},\ }\href@noop {} {\bibfield  {journal} {\bibinfo
  {journal} {Phys. Rev. Appl.}\ }\textbf {\bibinfo {volume} {10}},\ \bibinfo
  {pages} {034059} (\bibinfo {year} {2018})}\BibitemShut {NoStop}%
\bibitem [{\citenamefont {Hishinuma}\ \emph {et~al.}(2001)\citenamefont
  {Hishinuma}, \citenamefont {Geballe}, \citenamefont {Moyzhes},\ and\
  \citenamefont {Kenny}}]{hishinuma2001refrigeration}%
  \BibitemOpen
  \bibfield  {author} {\bibinfo {author} {\bibfnamefont {Y.}~\bibnamefont
  {Hishinuma}}, \bibinfo {author} {\bibfnamefont {T.}~\bibnamefont {Geballe}},
  \bibinfo {author} {\bibfnamefont {B.}~\bibnamefont {Moyzhes}},\ and\ \bibinfo
  {author} {\bibfnamefont {T.~W.}\ \bibnamefont {Kenny}},\ }\bibfield  {title}
  {\bibinfo {title} {Refrigeration by combined tunneling and thermionic
  emission in vacuum: Use of nanometer scale design},\ }\href@noop {}
  {\bibfield  {journal} {\bibinfo  {journal} {Appl. Phys. Lett.}\ }\textbf
  {\bibinfo {volume} {78}},\ \bibinfo {pages} {2572} (\bibinfo {year}
  {2001})}\BibitemShut {NoStop}%
\bibitem [{\citenamefont {Rodriguez-Nieva}\ \emph {et~al.}(2016)\citenamefont
  {Rodriguez-Nieva}, \citenamefont {Dresselhaus},\ and\ \citenamefont
  {Song}}]{rodriguez2016enhanced}%
  \BibitemOpen
  \bibfield  {author} {\bibinfo {author} {\bibfnamefont {J.~F.}\ \bibnamefont
  {Rodriguez-Nieva}}, \bibinfo {author} {\bibfnamefont {M.~S.}\ \bibnamefont
  {Dresselhaus}},\ and\ \bibinfo {author} {\bibfnamefont {J.~C.}\ \bibnamefont
  {Song}},\ }\bibfield  {title} {\bibinfo {title} {Enhanced
  thermionic-dominated photoresponse in graphene schottky junctions},\
  }\href@noop {} {\bibfield  {journal} {\bibinfo  {journal} {Nano Lett.}\
  }\textbf {\bibinfo {volume} {16}},\ \bibinfo {pages} {6036} (\bibinfo {year}
  {2016})}\BibitemShut {NoStop}%
\bibitem [{\citenamefont {Misra}\ \emph {et~al.}(2017)\citenamefont {Misra},
  \citenamefont {Upadhyay~Kahaly},\ and\ \citenamefont
  {Mishra}}]{misra2017thermionic}%
  \BibitemOpen
  \bibfield  {author} {\bibinfo {author} {\bibfnamefont {S.}~\bibnamefont
  {Misra}}, \bibinfo {author} {\bibfnamefont {M.}~\bibnamefont
  {Upadhyay~Kahaly}},\ and\ \bibinfo {author} {\bibfnamefont {S.}~\bibnamefont
  {Mishra}},\ }\bibfield  {title} {\bibinfo {title} {Thermionic emission from
  monolayer graphene, sheath formation and its feasibility towards thermionic
  converters},\ }\href@noop {} {\bibfield  {journal} {\bibinfo  {journal} {J.
  Appl. Phys.}\ }\textbf {\bibinfo {volume} {121}},\ \bibinfo {pages} {065102}
  (\bibinfo {year} {2017})}\BibitemShut {NoStop}%
\bibitem [{\citenamefont {Sinha}\ and\ \citenamefont
  {Lee}(2014)}]{sinha2014ideal}%
  \BibitemOpen
  \bibfield  {author} {\bibinfo {author} {\bibfnamefont {D.}~\bibnamefont
  {Sinha}}\ and\ \bibinfo {author} {\bibfnamefont {J.~U.}\ \bibnamefont
  {Lee}},\ }\bibfield  {title} {\bibinfo {title} {Ideal graphene/silicon
  schottky junction diodes},\ }\href@noop {} {\bibfield  {journal} {\bibinfo
  {journal} {Nano Lett.}\ }\textbf {\bibinfo {volume} {14}},\ \bibinfo {pages}
  {4660} (\bibinfo {year} {2014})}\BibitemShut {NoStop}%
\bibitem [{\citenamefont {Ang}\ \emph {et~al.}(2018)\citenamefont {Ang},
  \citenamefont {Yang},\ and\ \citenamefont {Ang}}]{ang2018universal}%
  \BibitemOpen
  \bibfield  {author} {\bibinfo {author} {\bibfnamefont {Y.~S.}\ \bibnamefont
  {Ang}}, \bibinfo {author} {\bibfnamefont {H.~Y.}\ \bibnamefont {Yang}},\ and\
  \bibinfo {author} {\bibfnamefont {L.}~\bibnamefont {Ang}},\ }\bibfield
  {title} {\bibinfo {title} {Universal scaling laws in schottky
  heterostructures based on two-dimensional materials},\ }\href@noop {}
  {\bibfield  {journal} {\bibinfo  {journal} {Phys. Rev. Lett.}\ }\textbf
  {\bibinfo {volume} {121}},\ \bibinfo {pages} {056802} (\bibinfo {year}
  {2018})}\BibitemShut {NoStop}%
\bibitem [{\citenamefont {Trushin}(2018{\natexlab{a}})}]{trushin2018theory1}%
  \BibitemOpen
  \bibfield  {author} {\bibinfo {author} {\bibfnamefont {M.}~\bibnamefont
  {Trushin}},\ }\bibfield  {title} {\bibinfo {title} {Theory of thermionic
  emission from a two-dimensional conductor and its application to a
  graphene-semiconductor schottky junction},\ }\href@noop {} {\bibfield
  {journal} {\bibinfo  {journal} {Appl. Phys. Lett.}\ }\textbf {\bibinfo
  {volume} {112}},\ \bibinfo {pages} {171109} (\bibinfo {year}
  {2018}{\natexlab{a}})}\BibitemShut {NoStop}%
\bibitem [{\citenamefont {Trushin}(2018{\natexlab{b}})}]{trushin2018theory2}%
  \BibitemOpen
  \bibfield  {author} {\bibinfo {author} {\bibfnamefont {M.}~\bibnamefont
  {Trushin}},\ }\bibfield  {title} {\bibinfo {title} {Theory of photoexcited
  and thermionic emission across a two-dimensional graphene-semiconductor
  schottky junction},\ }\href@noop {} {\bibfield  {journal} {\bibinfo
  {journal} {Phys. Rev. B}\ }\textbf {\bibinfo {volume} {97}},\ \bibinfo
  {pages} {195447} (\bibinfo {year} {2018}{\natexlab{b}})}\BibitemShut
  {NoStop}%
\bibitem [{\citenamefont {Neto}\ \emph {et~al.}(2009)\citenamefont {Neto},
  \citenamefont {Guinea}, \citenamefont {Peres}, \citenamefont {Novoselov},\
  and\ \citenamefont {Geim}}]{neto2009electronic}%
  \BibitemOpen
  \bibfield  {author} {\bibinfo {author} {\bibfnamefont {A.~C.}\ \bibnamefont
  {Neto}}, \bibinfo {author} {\bibfnamefont {F.}~\bibnamefont {Guinea}},
  \bibinfo {author} {\bibfnamefont {N.~M.}\ \bibnamefont {Peres}}, \bibinfo
  {author} {\bibfnamefont {K.~S.}\ \bibnamefont {Novoselov}},\ and\ \bibinfo
  {author} {\bibfnamefont {A.~K.}\ \bibnamefont {Geim}},\ }\bibfield  {title}
  {\bibinfo {title} {The electronic properties of graphene},\ }\href@noop {}
  {\bibfield  {journal} {\bibinfo  {journal} {Rev. Mod. Phys.}\ }\textbf
  {\bibinfo {volume} {81}},\ \bibinfo {pages} {109} (\bibinfo {year}
  {2009})}\BibitemShut {NoStop}%
\bibitem [{\citenamefont {Ang}\ \emph {et~al.}(2019)\citenamefont {Ang},
  \citenamefont {Chen}, \citenamefont {Tan},\ and\ \citenamefont
  {Ang}}]{ang2019generalized}%
  \BibitemOpen
  \bibfield  {author} {\bibinfo {author} {\bibfnamefont {Y.~S.}\ \bibnamefont
  {Ang}}, \bibinfo {author} {\bibfnamefont {Y.}~\bibnamefont {Chen}}, \bibinfo
  {author} {\bibfnamefont {C.}~\bibnamefont {Tan}},\ and\ \bibinfo {author}
  {\bibfnamefont {L.}~\bibnamefont {Ang}},\ }\bibfield  {title} {\bibinfo
  {title} {Generalized high-energy thermionic electron injection at graphene
  interface},\ }\href@noop {} {\bibfield  {journal} {\bibinfo  {journal} {Phys.
  Rev. Appl.}\ }\textbf {\bibinfo {volume} {12}},\ \bibinfo {pages} {014057}
  (\bibinfo {year} {2019})}\BibitemShut {NoStop}%
\bibitem [{\citenamefont {Javadi}\ \emph {et~al.}(2020)\citenamefont {Javadi},
  \citenamefont {Noroozi},\ and\ \citenamefont {Abdi}}]{javadi2020kinetics}%
  \BibitemOpen
  \bibfield  {author} {\bibinfo {author} {\bibfnamefont {M.}~\bibnamefont
  {Javadi}}, \bibinfo {author} {\bibfnamefont {A.}~\bibnamefont {Noroozi}},\
  and\ \bibinfo {author} {\bibfnamefont {Y.}~\bibnamefont {Abdi}},\ }\bibfield
  {title} {\bibinfo {title} {Kinetics of charge carriers across a
  graphene-silicon schottky junction},\ }\href@noop {} {\bibfield  {journal}
  {\bibinfo  {journal} {Phys. Rev. Appl.}\ }\textbf {\bibinfo {volume} {14}},\
  \bibinfo {pages} {064048} (\bibinfo {year} {2020})}\BibitemShut {NoStop}%
\bibitem [{\citenamefont {Gueymard}\ \emph {et~al.}(2002)\citenamefont
  {Gueymard}, \citenamefont {Myers},\ and\ \citenamefont
  {Emery}}]{gueymard2002proposed}%
  \BibitemOpen
  \bibfield  {author} {\bibinfo {author} {\bibfnamefont {C.~A.}\ \bibnamefont
  {Gueymard}}, \bibinfo {author} {\bibfnamefont {D.}~\bibnamefont {Myers}},\
  and\ \bibinfo {author} {\bibfnamefont {K.}~\bibnamefont {Emery}},\ }\bibfield
   {title} {\bibinfo {title} {Proposed reference irradiance spectra for solar
  energy systems testing},\ }\href@noop {} {\bibfield  {journal} {\bibinfo
  {journal} {Solar energy}\ }\textbf {\bibinfo {volume} {73}},\ \bibinfo
  {pages} {443} (\bibinfo {year} {2002})}\BibitemShut {NoStop}%
\bibitem [{\citenamefont {Liang}\ \emph {et~al.}(2017)\citenamefont {Liang},
  \citenamefont {Liu}, \citenamefont {Hu}, \citenamefont {Zhou},\ and\
  \citenamefont {Ang}}]{liang2017thermionic}%
  \BibitemOpen
  \bibfield  {author} {\bibinfo {author} {\bibfnamefont {S.-J.}\ \bibnamefont
  {Liang}}, \bibinfo {author} {\bibfnamefont {B.}~\bibnamefont {Liu}}, \bibinfo
  {author} {\bibfnamefont {W.}~\bibnamefont {Hu}}, \bibinfo {author}
  {\bibfnamefont {K.}~\bibnamefont {Zhou}},\ and\ \bibinfo {author}
  {\bibfnamefont {L.}~\bibnamefont {Ang}},\ }\bibfield  {title} {\bibinfo
  {title} {Thermionic energy conversion based on graphene van der waals
  heterostructures},\ }\href@noop {} {\bibfield  {journal} {\bibinfo  {journal}
  {Sci. Rep.}\ }\textbf {\bibinfo {volume} {7}},\ \bibinfo {pages} {1}
  (\bibinfo {year} {2017})}\BibitemShut {NoStop}%
\bibitem [{\citenamefont {Trucchi}\ \emph {et~al.}(2018)\citenamefont
  {Trucchi}, \citenamefont {Bellucci}, \citenamefont {Girolami}, \citenamefont
  {Calvani}, \citenamefont {Cappelli}, \citenamefont {Orlando}, \citenamefont
  {Polini}, \citenamefont {Silvestroni}, \citenamefont {Sciti},\ and\
  \citenamefont {Kribus}}]{trucchi2018solar}%
  \BibitemOpen
  \bibfield  {author} {\bibinfo {author} {\bibfnamefont {D.~M.}\ \bibnamefont
  {Trucchi}}, \bibinfo {author} {\bibfnamefont {A.}~\bibnamefont {Bellucci}},
  \bibinfo {author} {\bibfnamefont {M.}~\bibnamefont {Girolami}}, \bibinfo
  {author} {\bibfnamefont {P.}~\bibnamefont {Calvani}}, \bibinfo {author}
  {\bibfnamefont {E.}~\bibnamefont {Cappelli}}, \bibinfo {author}
  {\bibfnamefont {S.}~\bibnamefont {Orlando}}, \bibinfo {author} {\bibfnamefont
  {R.}~\bibnamefont {Polini}}, \bibinfo {author} {\bibfnamefont
  {L.}~\bibnamefont {Silvestroni}}, \bibinfo {author} {\bibfnamefont
  {D.}~\bibnamefont {Sciti}},\ and\ \bibinfo {author} {\bibfnamefont
  {A.}~\bibnamefont {Kribus}},\ }\bibfield  {title} {\bibinfo {title} {Solar
  thermionic-thermoelectric generator (st2g): Concept, materials engineering,
  and prototype demonstration},\ }\href@noop {} {\bibfield  {journal} {\bibinfo
   {journal} {Adv. Energy Mater.}\ }\textbf {\bibinfo {volume} {8}},\ \bibinfo
  {pages} {1802310} (\bibinfo {year} {2018})}\BibitemShut {NoStop}%
\bibitem [{\citenamefont {Freitag}\ \emph {et~al.}(2010)\citenamefont
  {Freitag}, \citenamefont {Chiu}, \citenamefont {Steiner}, \citenamefont
  {Perebeinos},\ and\ \citenamefont {Avouris}}]{freitag2010thermal}%
  \BibitemOpen
  \bibfield  {author} {\bibinfo {author} {\bibfnamefont {M.}~\bibnamefont
  {Freitag}}, \bibinfo {author} {\bibfnamefont {H.-Y.}\ \bibnamefont {Chiu}},
  \bibinfo {author} {\bibfnamefont {M.}~\bibnamefont {Steiner}}, \bibinfo
  {author} {\bibfnamefont {V.}~\bibnamefont {Perebeinos}},\ and\ \bibinfo
  {author} {\bibfnamefont {P.}~\bibnamefont {Avouris}},\ }\bibfield  {title}
  {\bibinfo {title} {Thermal infrared emission from biased graphene},\
  }\href@noop {} {\bibfield  {journal} {\bibinfo  {journal} {Nat.
  Nanotechnology}\ }\textbf {\bibinfo {volume} {5}},\ \bibinfo {pages} {497}
  (\bibinfo {year} {2010})}\BibitemShut {NoStop}%
\bibitem [{\citenamefont {Rahman}\ and\ \citenamefont
  {Nojeh}(2020)}]{rahman2020interplay}%
  \BibitemOpen
  \bibfield  {author} {\bibinfo {author} {\bibfnamefont {E.}~\bibnamefont
  {Rahman}}\ and\ \bibinfo {author} {\bibfnamefont {A.}~\bibnamefont {Nojeh}},\
  }\bibfield  {title} {\bibinfo {title} {Interplay between near-field radiative
  coupling and space-charge effects in a microgap thermionic energy converter
  under fixed heat input},\ }\href@noop {} {\bibfield  {journal} {\bibinfo
  {journal} {Physical Review Applied}\ }\textbf {\bibinfo {volume} {14}},\
  \bibinfo {pages} {024082} (\bibinfo {year} {2020})}\BibitemShut {NoStop}%
\bibitem [{\citenamefont {Massicotte}\ \emph {et~al.}(2016)\citenamefont
  {Massicotte}, \citenamefont {Schmidt}, \citenamefont {Vialla}, \citenamefont
  {Watanabe}, \citenamefont {Taniguchi}, \citenamefont {Tielrooij},\ and\
  \citenamefont {Koppens}}]{massicotte2016photo}%
  \BibitemOpen
  \bibfield  {author} {\bibinfo {author} {\bibfnamefont {M.}~\bibnamefont
  {Massicotte}}, \bibinfo {author} {\bibfnamefont {P.}~\bibnamefont {Schmidt}},
  \bibinfo {author} {\bibfnamefont {F.}~\bibnamefont {Vialla}}, \bibinfo
  {author} {\bibfnamefont {K.}~\bibnamefont {Watanabe}}, \bibinfo {author}
  {\bibfnamefont {T.}~\bibnamefont {Taniguchi}}, \bibinfo {author}
  {\bibfnamefont {K.-J.}\ \bibnamefont {Tielrooij}},\ and\ \bibinfo {author}
  {\bibfnamefont {F.~H.}\ \bibnamefont {Koppens}},\ }\bibfield  {title}
  {\bibinfo {title} {Photo-thermionic effect in vertical graphene
  heterostructures},\ }\href@noop {} {\bibfield  {journal} {\bibinfo  {journal}
  {Nat. Commun.}\ }\textbf {\bibinfo {volume} {7}},\ \bibinfo {pages} {1}
  (\bibinfo {year} {2016})}\BibitemShut {NoStop}%
\bibitem [{\citenamefont {Rouklove}(1966)}]{rouklove1966thermionic}%
  \BibitemOpen
  \bibfield  {author} {\bibinfo {author} {\bibfnamefont {P.}~\bibnamefont
  {Rouklove}},\ }\bibfield  {title} {\bibinfo {title} {Thermionic converter and
  generator development},\ }\href@noop {} {\bibfield  {journal} {\bibinfo
  {journal} {Space Programs;(United States)}\ }\textbf {\bibinfo {volume} {4}}
  (\bibinfo {year} {1966})}\BibitemShut {NoStop}%
\end{thebibliography}%

\end{document}